\documentclass[%
 amsmath,amssymb,
 aps,
 twocolumn,
 showpacs,
 superscriptaddress,
 groupedaddress,
 prb
]{revtex4}

\usepackage{amsmath}
\usepackage{amssymb}
\usepackage{float}
\usepackage{epsfig}
\usepackage[utf8]{inputenc}
\usepackage[T1]{fontenc}
\usepackage{lmodern}
\usepackage{multirow}
\usepackage{diagbox}
\usepackage{graphicx}
\usepackage{dcolumn}
\usepackage{bm}
\usepackage{hyperref}
\usepackage[mathlines]{lineno}
\usepackage{xcolor}
\usepackage{soul}
\usepackage[english]{babel}
\makeatletter\AtBeginDocument{\let\@elt\relax}\makeatother

\def\<{\langle}
\def\>{\rangle}

\def\beq{\begin{equation}}
\def\eeq{\end{equation}}

\newcommand{\bea}{\begin{eqnarray}}
\newcommand{\eea}{\end{eqnarray}}

\newcommand{\tdos}{$\mathrm{TDOS~}$}

\DeclareUnicodeCharacter{0308}{480}

\def\lsim{\mathrel{\rlap{\lower4pt\hbox{\hskip1pt$\sim$}}
    \raise1pt\hbox{$<$}}}         
\def\gsim{\mathrel{\rlap{\lower4pt\hbox{\hskip1pt$\sim$}}
    \raise1pt\hbox{$>$}}}         

\usepackage{url}
\usepackage{amsfonts}
\usepackage{textcomp}
\def\BibTeX{{\rm B\kern-.05em{\sc i\kern-.025em b}\kern-.08em
    T\kern-.1667em\lower.7ex\hbox{E}\kern-.125emX}}

\begin{document}

\title{ Tunneling density of states in Luttinger Liquid in proximity to a superconductor: Effect of non-local interaction }

\author{Amulya Ratnakar  and Sourin Das \\
{\emph{Department of Physical Sciences,\\
Indian Institute of Science Education and Research (IISER) Kolkata \\
Mohanpur - 741 246,
West Bengal, India}}} 

\date{\today}
                              
\begin{abstract}
A recent study have shown that it is possible to have enhancement, in contrast to an expected suppression, in tunneling density of states (TDOS) in a Luttinger liquid (LL) which is solely driven by the non-local density-density interactions. Also, it is well known that a LL in proximity to a superconductor (SC) shows enhancement in TDOS in the vicinity of the junction in the zero energy limit. In this paper, we study the interplay of nonlocal density-density interaction and superconducting correlations in the TDOS in the vicinity of the SC-LL junction, where the LL maybe realized on the edge of an integer or a fractional quantum Hall state. We show that the interplay of superconducting proximity effect and non-local interactions can give rise to enhancement in TDOS in the weak interaction limit, beyond what was previously observed. We also show that, in the full parameter regime comprising both, the local and the non-local interaction,  the region of enhanced  TDOS for LL junction with "superconducting" boundary condition and that of  "non-superconducting charge conserving" boundary condition (discussed in Phys. Rev. B 104, 045402 (2021)) are mutually exclusive. We show that this fact can be understood in terms of symmetry relation established between the superconducting and non-superconducting sectors of the theory. We compare the dependence of the proximity induced pair potential and TDOS as a function of distance $x$ from the junction. We demonstrate that the dependence of the spatial power law exponent for the `TDOS($x$)/TDOS($x\rightarrow 0$)' and the `pair-potential($x$)/pair-potential($x\rightarrow 0$)' are distinct function of the various local and non-local interaction parameters, which implies that the TDOS enhancement can not be directly attributed to the proximity induced pair potential in the LL. 

\end{abstract}                              
                              
\maketitle

\section{Introduction}
The low energy physics of an interacting one dimensional (1-D) electronic system is described by the universal Luttinger liquid (LL) model which exhibits "non-Fermi-liquid behavior" due to the absence of electronlike quasi particles in the low energy excitation spectrum
~\cite{Haldane_1981,kane1992transport,Srao2002,Tgiamarchi2004,von1998bosonization,oreg1996enhancement, goglin_reply, Kane1992Transmission, winkelholz1996anomalous, fradkin1999exploring, eggert2000scanning,fisher1997transport,aristov2010tunneling,maslov2005,DDVu}. Tunneling experiments have played a key role in probing and understanding the "non-Fermi liquid" properties of both, the chiral\cite{chang1996,chang_review,grayson1998,chang2002,milliken1996,Fabio_2DEG_C_ll,Grayson_2001_2DEG_C_ll,kang2000_2DEG_C_ll,chang2012_2DEG_C_ll,meier2014_2DEG_C_LL,turley1998_2DEG_C_ll,roddaro2004_2DEG_C_ll,neto2006,hu2012_graphene_C_ll,li2013_graphene,moreau2021_graphene_C_ll} and the non-chiral LLs~\cite{Dimitry_wire_LL,ning2014_wire_ll,sato2019_wire_ll,zhao2018_wire_ll,Aleshin2004_wire_ll,aleshin2005_wire_ll,zaitsev2000_wire_ll,bockrath1999_wire_ll}. For non-chiral LLs, such as 1-D quantum wires (QWs), it is difficult to observe LL behavior in a tunneling experimental setup because any residual disorder can affect the power law characteristic of the tunneling conductance. However, a chiral LL, realized at the edge of the FQH system, is immune to disorders and impurities at the boundary and as such, shows the characteristic power law behavior for tunneling conductance. To this end, experiments including  local electron tunneling between the edge states of a fractional quantum Hall (FQH) system realized on the gated 2 DEGs~\cite{chang1996,chang_review,grayson1998,chang2002,milliken1996,Fabio_2DEG_C_ll,Grayson_2001_2DEG_C_ll,kang2000_2DEG_C_ll,chang2012_2DEG_C_ll,meier2014_2DEG_C_LL,turley1998_2DEG_C_ll,roddaro2004_2DEG_C_ll} or between the Fermi-liquid (FL) lead and FQH edge realized on a graphene sheet~\cite{neto2006,hu2012_graphene_C_ll,li2013_graphene,moreau2021_graphene_C_ll} have been performed. The tunneling current is shown to have a power law suppression at the zero-bias limit, which provides the robust evidence of the LL behavior of the chiral edge modes of FQH system.

An exception to the suppression, leading to enhancement in TDOS, is predicted at the junction of multiple LLs.
~\cite{agarwal2009enhancement,Affleck2006junctions,ChamonJunction_y_junction,shi2016_y_junction,Aristov_wolfie_y_junction,aggarwal2015enhancement_y_junction,aristov2013chiral_y_junction,SDas_manu_2020tunneling_y_junction,agarwal2010power}. 
Due to strong correlation which are localized  at the junction of multiple LLs, fixed point (FP) structure has a possibility to host hole current which gets reflected in the LL in response to an incident electron current, which in turn was attributed as the reason for enhancement in TDOS at the junction.~\cite{agarwal2009enhancement}. 
Such fixed points showing enhancement in TDOS were found to be unstable against perturbation which can be switched on at the junction in the renormalization group (RG) sense, rendering it hard for such FPs to be observed in an experimental setup. A recent study done by current authors, showed that for a minimal model of a junction of two LLs, it is possible to have simultaneous TDOS enhancement and stability at the junction, even in the absence of Andreev like process, if we switch on the nonlocal density-density interactions between the two LLs.
~\cite{Amulya2021}.

An independent scenario, which naturally supports the hole current in response to an incident electron current and TDOS enhancement, is that of a LL in proximity to a superconductor (SC). Previous studies have shown that for a junction between half wire LL and an  SC, in the strong coupling limit, TDOS shows enhancement in the vicinity of the junction~\cite{winkelholz1996anomalous}. Later, the study of such junctions was extended to the case of a junction of two or more LL half wires, where the junction itself was considered to be superconducting~\cite{Duality}. Duality relation were established between scaling dimension of various perturbation belonging to the current conserving (Normal) and current non-conserving (superconducting) fixed points of the theory and junction conductance were calculated, though the question of TDOS enhancement in the new scenario stayed unexplored.

Hence, it is pertinent to explore the effect of superconducting proximity effect on TDOS for a junction of two LL with a superconductor in the presence of non-local density-density interactions. It is assumed that the distance between the two LLs at the junction is less than the superconducting coherence length, such that the cross Andreev tunneling across the junction is facilitated. It may be expected for TDOS in the vicinity of the junction to show amplified enhancement, upon introducing the superconducting correlation at the junction, in the region of parameter space of bulk interaction in which the TDOS is already enhanced for non-superconducting current conserving (normal) fixed point~\cite{Amulya2021}. On the contrary, we find that the superconducting correlation induces suppression in this parameter regime. The parameter regime in which TDOS shows enhancement for both, Normal and superconducting junction, are mutually exclusive. This can be understood in terms of symmetry relations between the charge conserving and superconducting fixed point, as shown in Sec~\ref{4} and ~\ref{5}.
We report the possibility of enhancement in TDOS in a LL QW, beyond what was observed in Ref.~\cite{winkelholz1996anomalous}, in the presence of appropriate nonlocal density-density interactions, with superconducting boundary condition as the fixed point of theory. We also study the spatial dependencies of proximity induced pair correlation function and the TDOS. One could naively think if the enhancement of TDOS at the SC-LL junction is the consequence of proximity effect then the decay profile of both, the TDOS and the induced pair correlation function, to be the same. It was shown in the Ref.\cite{winkelholz1996anomalous}, that in the presence of local interactions, the induced pair amplitude decays faster as compared to the enhanced TDOS away from the SC-LL junction. We present a clear understanding of this mismatch in the decay  profile in terms of the Bogoliubov modes of the system and show that this behavior remains true even in the presence of non-local interaction.

This paper is organized in the following way: Section~\ref{section 1} reviews the results of TDOS, induced pair amplitude function and stability of the fixed point for a junction of single and double LL QWs in proximity to a superconductor (as shown in fig.~\ref{Fig. SC-LL}). Subsequent sections deals with the more general case of a junction of edge state of two factional (or integer) quantum Hall (QH) system in proximity to a superconductor with all possible density-density interaction between them. Section~\ref{3} primarily deals with the diagonalization of the interacting edge Hamiltonian by finding the appropriate Bogoliubov modes for the system. A general framework to calculate the scaling dimension of perturbation operators and power law of different correlation function is presented. The results of TDOS enhancement and the stability scenario for the fixed point at the junction is analyzed in Sec.~\ref{4} and Sec.~\ref{5} for the case when the filling fraction for both the QH system is same and different respectively. We conclude the results and discuss the experimental relevance of our set-up in Sec.~\ref{6}.

\section{Review: SC-LL and LL-SC-LL junction} \label{section 1}
  
\subsection{SC-LL junction}
 
\begin{figure*}[t]
\centering
\includegraphics[scale=0.15]{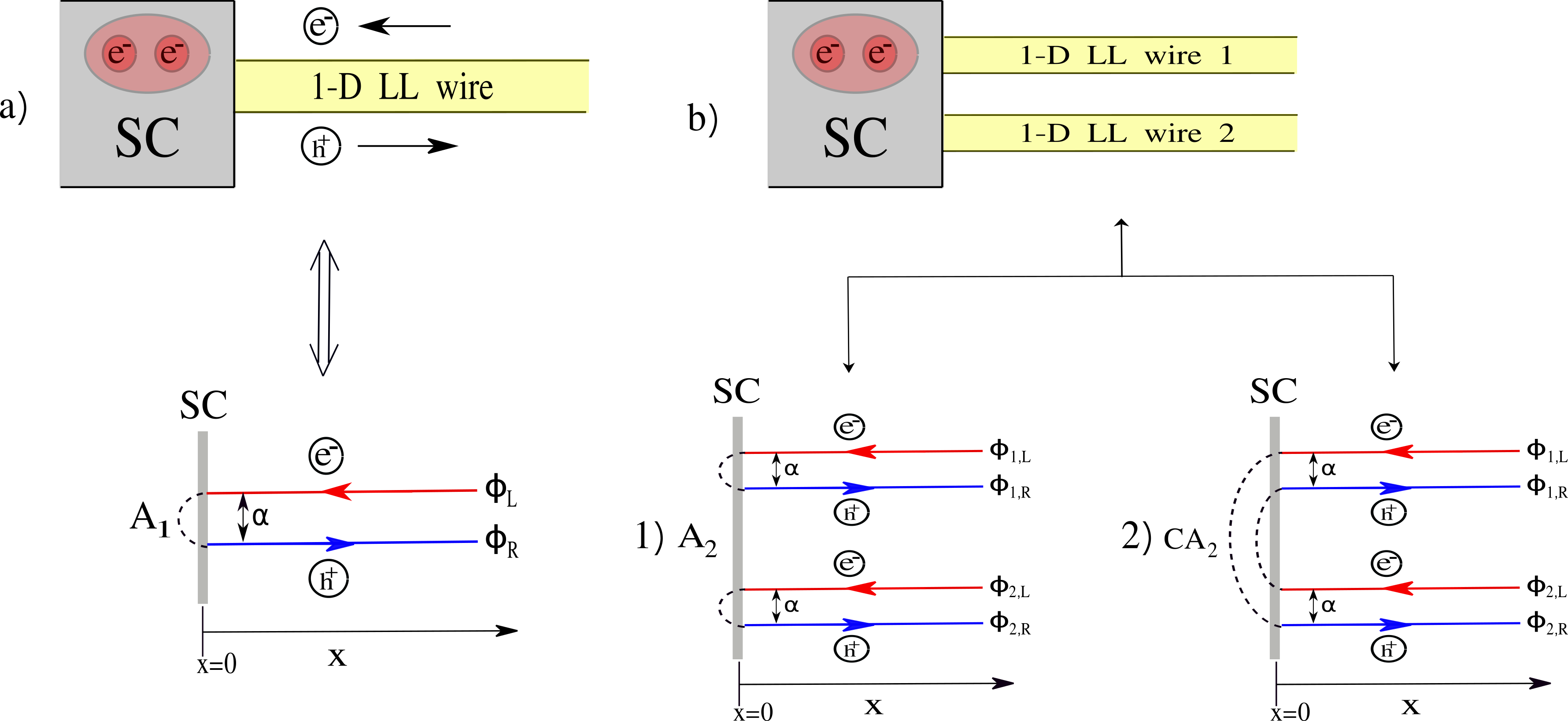}
\caption{Figure a) and b) corresponds to the physical setting of the LL QWs. Schematic figure shows the junction of superconductor with a) a single LL quantum wire (SC-LL junction) and b) two LL quantum wire (LL-SC-LL junction),  such that the distance between the two LL QWs at the junction is less than the superconducting coherence length, which can facilitate the cross Andreev tunneling at the junction. $\alpha$ is the inter-electron density-density interaction. The dashed line is indicative of the Andreev reflection process at the junction. Figure a) shows the SC-LL junction tuned to $\mathrm{A_{1}}$ fixed point. Figure b) shows the two possible fixed point, 1) corresponding to the weak coupling limit, in which junction is tuned to disconnected $\mathrm{A_{2}}$ fixed point (Note that $\mathrm{A_{2}}$ fixed point is the copy of two $\mathrm{A_{1}}$ fixed points)  and 2) corresponding to the strong coupling limit, in which junction is tuned to cross Andreev reflection $\mathrm{CA_{2}}$ fixed point.}
\label{Fig. SC-LL}
\end{figure*}

Consider the case of a spinless LL quantum wire (QW) strongly coupled to a superconductor at $x=0$ as shown in fig.~\ref{Fig. SC-LL}a. The fermionic field can be decomposed in terms of right (R) and left (L) moving components. One can use bosonization to define fermionic fields $\psi_{R/L}(x)$ in terms of bosonic fields $\phi_{R/L}(x)$ through the relation $\psi_{R/L}(x) = (F_{R/L}/\sqrt{2\pi \delta}) Exp\left[ i \phi_{R/L}(x)\right] $~\cite{Haldane_1981,Tgiamarchi2004,Srao2002,von1998bosonization}, where $F_{R/L}$ is the Klein factor corresponding to right/left moving chiral fields. Then the bosonized interacting Hamiltonian for the LL QW is given by

\begin{equation}
H = \hbar \pi v_{F}\int_{0}^{\infty} dx \left( \rho_{L}(x)^{2} + \rho_{R}(x)^{2} + 2\alpha \rho_{L}(x)\rho_{R}(x) \right)
\label{Eq: Single LL hamiltonian}.
\end{equation}

The electron density operators and electron current operators are given by $\rho_{R/L}(x,t) = \pm \left(1/2\pi\right) \partial_{x}\phi_{R/L}(x,t)$ and $j_{R/L}(x,t) = \pm(v_{F}/2\pi)\partial_{x}\phi_{R/L}(x,t)$ respectively~\cite{agarwal2009enhancement,Duality,Affleck2006junctions}. $\alpha$ is the local density-density interaction parameter between the `right' and `left' moving fields. The interacting bosonic fields $\phi_{R/L}$ can be expressed in terms of Bogoliubov (Bg) fields $\tilde{\phi}_{R/L}$ as 

\begin{equation}
\phi_{R/L}(x) = \frac{1}{2\sqrt{g}}\left( (g+1)\tilde{\phi}_{R/L}(x) -(g-1)\tilde{\phi}_{L/R}(x) \right) 
\end{equation}

where $g = \sqrt{1-\alpha}/\sqrt{1+\alpha}$ and $g<1$ ($g>1$) for repulsive (attractive) inter-electron interaction. In this paper, only the repulsive inter-electron interaction is considered, unless otherwise mentioned. The boundary condition (BC) that a left moving electronic current is reflected back as a right moving hole current  at the junction, $j_{R}(0) = -j_{L}(0)$~\cite{Affleck2006junctions,Duality}, is equivalent to $\phi_{R}(0) = -\phi_{L}(0) + C$ (where $C$ is an integration constant). Since the evaluation of scaling dimensions of operators around any fixed point does not depend on $C$, we can ignore it by taking $C=0$. The condition, $\phi_{R}(0)=-\phi_{L}(0)$, defines Andreev fixed point $(A_{1})$  at the junction. The BC on interacting fields gives the BC on the Bg fields as $\tilde{\phi}_{R}(0) = -\tilde{\phi}_{L}(0)$, which, at finite $x$, translates to $\tilde{\phi}_{R}(x) = -\tilde{\phi}_{L}(-x)$. 

The local TDOS~\cite{agarwal2009enhancement,kane1992transport} at energy $E$ and at finite distance $x$ away from the junction is given by

\begin{eqnarray}
\rho(x,E) &=& \int^{\infty}_{-\infty} \langle 0\vert \psi(x,t)\psi^{\dagger}(x,0)\vert0\rangle e^{-iEt} dt
\label{Eq: TDOS_def}
\end{eqnarray}

The TDOS is studied in the zero temperature limit ($T\rightarrow 0$). The Green function in Eq.~\ref{Eq: TDOS_def} is given by $G(x,t) = \langle \psi(x,t)\psi^{\dagger}(x,0) \rangle = \langle \psi_{R}(x,t)\psi_{R}^{\dagger}(x,0) \rangle + \langle \psi_{L}(x,t)\psi_{L}^{\dagger}(x,0) \rangle + e^{2ik_{F}x}\langle \psi_{R}(x,t)\psi_{L}^{\dagger}(x,0) \rangle + e^{-2ik_{F}x}\langle \psi_{L}(x,t)\psi_{R}^{\dagger}(x,0) \rangle$. In general, the oscillatory part goes to zero in the limit $L \rightarrow \infty$ and hence we focus on the non-oscillatory part which, in the limit $T \rightarrow 0$, is given by

\begin{eqnarray}
&&\langle \psi_{R}(x,t)\psi^{\dagger}_{R}(x,0) \rangle = \langle \psi_{L}(x,t)\psi^{\dagger}_{L}(x,0)\rangle \nonumber\\
&=& \frac{1}{2\pi \delta}\left( \frac{i\delta}{i\delta -vt} \right)^{\frac{1}{2}\left(g + \frac{1}{g} \right)} \left( \frac{(i\delta)^{2} - 4 x^{2}}{(i\delta - vt)^{2} - 4x^{2}}\right)^{\frac{\bar{a}}{4}\left(\frac{1}{g}-g\right)},\nonumber \\
\label{Eq:TDOS_correlation}
\end{eqnarray}

where $v$ is the renormalized velocity and is given by $v = v_{F}\sqrt{1-\alpha^{2}}$. $\delta$ is the short cutoff length such that $x>\delta$. In general, a fixed point is denoted by a current splitting matrix at the junction connecting multiple left moving bosonic fields to the multiple right moving bosonic fields at the junction. $\bar{a}$, then, is the diagonal element of the current splitting matrix, which in the case of a single LL in contact with a superconductor, is given by $\bar{a} = -1$. While calculating the Green's function $\langle \psi_{R/L}(x,t)\psi^{\dagger}_{R/L}(x,0) \rangle  = \langle e^{i\phi_{R/L}(x,t)}e^{-i\phi_{R/L}(x,0)}\rangle$, we must note that the contribution to the term $(i\delta/i\delta - vt)^{(1+g^{2})/2g}$ comes from the bosonic correlation function, $\langle \tilde{\phi}_{R/L}(x,t)\tilde{\phi}_{R/L}(x,0) \rangle$ and is insensitive to the BC at the junction. While on the other hand, the contribution to the term $((i\delta)^{2} - 4x^{2})/((i\delta - vt)^{2} -4x^{2}))^{\bar{a}(1-g^{2})/4g}$ in the Green's function comes from the correlation function between $\langle \tilde{\phi}_{R/L}(x,t)\tilde{\phi}_{L/R}(x,0) \rangle$ and depends on the BC at the junction because at finite $x$ the $\tilde{\phi}_{R}(x)$ and $\tilde{\phi}_{L}(-x)$ are related through appropriate BC on the  Bg fields. In the limit $ x \rightarrow \infty$, only the contribution from $\langle \tilde{\phi}_{R/L}(x,t)\tilde{\phi}_{R/L}(x,0) \rangle$ in the Green's function remains. As a result of which, the TDOS becomes insensitive to the BC and results in the usual power law suppression of the TDOS in the zero energy limit. On the other hand, in the proximity to the junction, that is, in the limit $x \rightarrow 0$, TDOS gets contribution from both, $\langle \phi_{R/L}(x,t)\phi_{R/L}(x,0) \rangle$ and $\langle \tilde{\phi}_{R/L}(x,t)\tilde{\phi}_{L/R}(x,0)\rangle$, in the Green's function and hence deviates from the TDOS power law corresponding to the bulk of the LL, i.e, $(1/2)(g+1/g)$. For a LL in contact with an SC ($\bar{a} = -1$), TDOS at the junction is given by $\rho (x \rightarrow 0,E) \propto E^{g - 1}$ and shows enhancement in the zero Energy limit in the repulsive interaction regime $(g < 1 )$. In contrast to the SC-LL junction, a LL half wire, with perfect backscattering normal fixed point ($\mathrm{N_{B}}$) at the junction $(\bar{a} = 1)$, the TDOS is given by $\rho (x \rightarrow 0,E) \propto E^{1/g - 1}$ and is more suppressed at the junction as compared to the bulk in the $E\rightarrow 0$ limit. Note that as far as the TDOS is concerned, there is a symmetry relation between the $\mathrm{A_{1}}$ and the $\mathrm{N_{B}}$ fixed point in the $\alpha \rightarrow -\alpha$ exchange and is given by
\begin{equation}
    \Delta^{0}_{A_{1}}(\alpha) = \Delta^{0}_{N_{B}}(-\alpha)
    \label{Eq:symmetry_SC-LL}
\end{equation}

From Eq.~\ref{Eq:symmetry_SC-LL} it is clear that as far as enhancement in TDOS for $A_{1}$ fixed point in the presence of repulsive ($\alpha >0$ or $g<1$) interaction is concerned, the identical enhancement can be achieved for $\mathrm{N_{B}}$ point but in the presence of attractive interaction of the same strength ($\alpha <0$ or $g>1$).

The stability of the $A_{1}$ fixed point is determined against the electron backscattering operator, $\psi_{R}^{\dagger}(0)\psi_{L}(0)$, which has a scaling dimension of $2g$\cite{Duality}. The $A_{1}$ fixed point is stable if the scaling dimension of the $\psi^{\dagger}_{R}(0)\psi_{L}(0)$ is greater than 1, that is, if $g > 1/2$. 

At finite distance $x$ from the junction, the TDOS varies as a power law of $x$ in the limit $x << v/E$ and is given by 
\begin{equation}
\rho(x,E) \propto (\delta^{2} + 4 x^{2})^{\frac{\bar{a}}{4}\left( \frac{1}{g} - g\right)} \omega^{\frac{1}{2}\left(g + \frac{1}{g} + \bar{a}\left(\frac{1}{g}-g \right)\right)}
\label{Eq: TDOS_x_NS}
\end{equation}   

Note that for the $A_{1}$ fixed point, the enhanced TDOS at the junction decreases as $x^{(1/2)(g-1/g)}$. For the half wire LL tuned to $N_{B}$ fixed point, TDOS increases with $x$ as $x^{-(1/2)(g-1/g)}$. 

As far as the enhancement of TDOS in the vicinity of a superconducting junction is concerned, one might be tempted to interpret it as the sole consequence of proximity effect and expect the decay profile of both, the enhanced TDOS and the induced pair amplitude to be the same. In order to understand the supposed interdependence of TDOS on induced pair amplitude (if it exists) in LL, we must have a look at the spatial dependence of the two. The spatial dependence of the induced pair amplitude is given by the pair correlation function $F(x) = \langle \psi_{R}(x,t^{+})\psi_{L}(x,t) \rangle$~\cite{maslov1996josephson,fazio1997properties,michelsen2020current,Yacoby2020induced}, which in the limit $L \rightarrow \infty$ , is given by

\begin{eqnarray}
F(x) &=& \langle \psi_{R}(x,t^{+})\psi_{L}(x,t)\rangle = \frac{1}{2\pi\delta}\langle e^{i\phi_{R}(x,t^{+})}e^{i\phi_{L}(x,t)}\rangle \nonumber\\
&=& \frac{1}{2\pi\delta}\left(i \delta\right)^{\frac{1}{g}}\left(2x + i \delta \right)^{\frac{-(1+g)}{2g}} \left(-2x + i\delta \right)^{\frac{-(1-g)}{2g}}
\label{Eq: PA_NS}
\end{eqnarray}

In the limit $x>> \delta$, the pair correlation function varies as  $F(x) \propto (1/x)^{1/g}$. Here, we introduce a term, 'relative TDOS' ('relative pair amplitude'), which is defined as the ratio of TDOS (pair amplitude) at finite distance x from the junction to the TDOS (pair amplitude) at the junction, that is, $\rho(x,E\rightarrow 0)/\rho(x\rightarrow 0,E\rightarrow 0)$ ($F(x)/F(x\rightarrow 0)$). Note that $\rho(x,E\rightarrow 0)/\rho(x\rightarrow 0,E\rightarrow 0)$ decays at a slower rate with $x$ as compared to $F(x)/F(x\rightarrow 0)$ function. As a result of which, the enhancement in TDOS persists over larger distances as compared to the decay length of the pair correlation function. Hence, the TDOS enhancement in the vicinity of a superconductor cannot be simply explained in terms of finite cooper pair density in the LL.

\begin{table*}[ht]
\centering
\begin{tabular}{|p{0.45\linewidth}|p{0.45\linewidth}|}
\hline
Current Conserving Normal junction & Superconducting junction \\
\hline
a) N=1

-Perfectly backscattering $\mathrm{N_{B}}$ fixed point with $\bar{a} = 1$, TDOS is suppressed for $g<1$. Note that, the fixed point is stable against the Andreev reflection operator for $g<1$\cite{Duality}.

& a) N=1 (LL-SC junction)
 
-Disconnected Andreev fixed point $A_{1}$, with $\bar{a} = -1$. TDOS is enhanced for $g<1$ and the $\mathrm{A_{1}}$ fixed point is stable against electron backscattering operator for $g>1/2$.\\
\hline
b) N=2

-Disconnected fixed point $\mathrm{DN_{2}} = \mathbb{I}_{2\times 2}$. TDOS is suppressed, and the fixed point is stable against electron tunneling operator for $g<1$.

-Connected fixed point. TDOS is suppressed, and the junction is unstable against electron backscattering operator for $g<1$.  
& b) N=2 (LL-SC-LL junction)

-Disconnected Andreev fixed point $\mathrm{A_{2}}$. TDOS is enhanced, but the junction is unstable against CAR and electron backscattering operator for $g<1$.

-Cross Andreev fixed point $\mathrm{CA_{2}}$. TDOS is suppressed, and the fixed point is unstable against electron backscattering operator for $g<1$.\\
\hline
\end{tabular}
\caption{The table provides results of TDOS enhancement and stability of the junction fixed point in absence of non-local interaction for current conserving boundary condition and their counterparts in the superconducting sector.}  
\label{Table:SC-LL} 
\end{table*}

\subsection{LL-SC-LL junction} 

Here we consider two LL QWs, namely wire 1 and wire 2,  in proximity to a superconductor as shown in fig.~\ref{Fig. SC-LL}b. Similar to Eq.~\ref{Eq: Single LL hamiltonian}, the bosonized Hamiltonian for the two LL QWs is given by

\begin{eqnarray}
H &=& \frac{\hbar v_{F}}{4\pi}\sum_{i=1}^{2}\int_{0}^{\infty} dx \left[\left(\partial_{x}\phi_{iL}(x)\right)^{2} + \left(\partial_{x}\phi_{iR}(x)\right)^{2} \right. \nonumber\\
 && \left.- 2\alpha \left(\partial_{x}\phi_{iL}(x)\right) \left(\partial_{x}\phi_{iR}(x)\right) \right] \nonumber
\end{eqnarray}
\begin{eqnarray}
\phi_{iR/L}(x) &=& \frac{1}{2\sqrt{g}}\left( (g+1)\tilde{\phi}_{iR/L}(x) -(g-1)\tilde{\phi}_{iL/R}(x) \right) \nonumber\\
\end{eqnarray}

There are four possible fixed point for such a junction: 
\begin{itemize}
\item[1)] Disconnected Normal ($\mathrm{DN_{2}}$) fixed point, where the QWs are disconnected from each other and also with the SC. The incident incoming electron at the junction reflects back as the outgoing electron in the same LL QW.

\item[2)] Fully transmitting charge conserving fixed point, where the QWs are strongly coupled to each other at the junction but disconnected from the superconductor, such that an incoming electron along the wire $1$ is perfectly transmitted in the wire $2$ and vice-versa.

\item[3)] Disconnected Andreev $(\mathrm{A_{2}})$ fixed point, where the QWs are disconnected with each other but are strongly coupled to the superconductor, such that the incident incoming electron current at the junction reflects as the outgoing hole current in the same LL wire. 

\item[4)] Cross Andreev reflection ($\mathrm{CA_{2}}$) fixed point, where the QWs are connected to the superconductor and strongly coupled to each other also. The incoming electron current along wire 1 is perfectly transmitted as a hole current in wire 2 and vice-versa\cite{bignon2004current_CAR,falci2001correlated_CAR,deutscher2000coupling_CAR,sadagashi2018cooper,sadagashi2019dominant,jelena2018}.

\end{itemize}

Here, we primarily focus on the Andreev fixed point $\mathrm{A_{2}}$ and the cross Andreev reflection $\mathrm{CA_{2}}$ fixed points, which are given by

\begin{equation}
A_{2} = \begin{pmatrix}
		-1 & 0 \\
		0 & -1 
		\end{pmatrix};\;\; \mathrm{CA_{2}} = \begin{pmatrix}
								 0 & -1 \\
								 -1 & 0	
								 \end{pmatrix}
\end{equation}

We first consider the case where the junction is tuned to $\mathrm{CA_{2}}$ fixed point. Since, in this case $\bar{a}=0$, from Eq.~\ref{Eq: TDOS_x_NS}, we note that the TDOS becomes independent of $`x'$ and behaves as a translation invariant single QW even when we have a superconducting junction at $x=0$. TDOS shows power law suppression, $\rho(x,E) \propto E^{(1/2)(g+1/g)}$, in the zero energy limit. This is counterintuitive in the light of previous studies, where we expect the TDOS to show enhancement in the vicinity of a superconductor. As it turns out, the fraction of incident electron current which reflects back as a hole solely decides the criteria of TDOS enhancement, and not just the mere presence of hole current at the junction.

The stability of the $\mathrm{CA_{2}}$ fixed point is determined against Andreev backscattering (AR) operator $(\psi_{iR}(0)\psi_{iL}(0))$, electron tunneling operator $(\psi^{\dagger}_{2R}(0)\psi_{1L}(0))$ and electron backscattering operator $(\psi^{\dagger}_{iR}(0)\psi_{iL}(0))$ at the junction. The scaling dimensions of these operators are $1/g$\cite{Duality}, $(g+1/g)$ and $g$ respectively. Hence, the $\mathrm{CA_{2}}$ fixed point is unstable against electron backscattering operator in the repulsive interaction regime $(g<1)$.  

Now we study the spatial dependence of induced pair correlation function. For a superconducting junction of two LL QWs tunes to $\mathrm{CA_{2}}$ fixed point, the induced pair correlation function is calculated between the right (R) and left (L) moving fields  of the different LL QWs and is termed as non-local pair correlation function, $F_{12}(x)$. The non-local induced pair correlations in the limit $L \rightarrow \infty$ is given by    

\begin{eqnarray}
F_{12}(x) &=& \langle \psi_{1R}(x,t^{+})\psi_{2L}(x,t)) \rangle = \frac{1}{2\pi \delta} \langle e^{i\phi_{1R}(x,t^{+})\phi_{2L}(x,t)} \rangle \nonumber\\
&=& \frac{1}{2\pi \delta}\left(i \delta\right)^{\frac{1}{2}\left(g + \frac{1}{g}\right)}\left( 2x + i \delta\right)^{-\frac{(1+g)^{2}}{4g}} \nonumber\\
&&\times \left(-2x +i\delta\right)^{-\frac{(1-g)^{2}}{4g}}
\label{Eq: PA_M2}  
\end{eqnarray} 

The induced pair amplitude is given by the real part of the $F_{12}(x)$, which, in the limit $x>>\delta $, decays as $F_{12}(x)\propto (1/x)^{(1/2)(g+1/g)}$ with distance $x$ from the junction. The power law decay of $F_{12}$ is independent of the type of the interaction present in the system (be it repulsive or attractive). The spatial dependence of both $\mathrm{F_{12}}(x)$ and TDOS are different, as the latter is independent of $x$.   

For the disconnected $A_{2}$ fixed point, the TDOS spatial and energy dependence remains the same as that of given in Eq.~\ref{Eq: TDOS_x_NS}. The TDOS is enhanced at the junction and decays as $x^{(1/2)(g-1/g)}$ away from the junction in both the wires. The local induced pair correlation function, $\langle\psi_{iR}^{\dagger}(x)\psi_{iL}^{\dagger}(x)\rangle$ is also the same as in Eq.~\ref{Eq: PA_NS} and decays as a power law of distance $x$ from the junction. 

The stability of the Andreev fixed point,  $A_{2}$ is determined against cross Andreev reflection (CAR) operator $\psi_{1R}(0)\psi_{2L}(0)$, electron backscattering operator $\psi^{\dagger}_{iR}(0)\psi_{iL}(0)$ and electron tunneling operator $\psi^{\dagger}_{1R}(0)\psi_{2L}(0)$ at the junction. The scaling dimensions of these operators are given by $g$, $2g$ and $g$ respectively. Hence, the $A_{2}$ fixed point is unstable against $\mathrm{CAR}$ and electron backscattering operator for $g<1$ and is stable against the electron backscattering operator for $g>1/2$. 

The results for the TDOS and the stability of the fixed point, as discussed in this section, are summarized in table~\ref{Table:SC-LL}. In this section, we reaffirmed that the enhancement in TDOS at the junction can not be solely attributed as the consequence of the induced pair amplitude as the spatial power law dependence of both are different, as pointed out in Ref.~\cite{winkelholz1996anomalous}. The pair correlation functions decays faster as compared to the enhanced TDOS at distance $x$ away from the junction, as a result of which TDOS remains enhanced up to distances larger than that of the decay length of induced pair amplitude. 

In subsequent sections, we introduce a more general system of a junction of QH edge states corresponding to two QH layers in proximity to a superconductor. We allow for non-local interaction between the two QH layer to exist and study the effect of both, the non-local interactions and superconducting boundary condition, on the TDOS and the stability of the junction fixed point. It would also be interesting to study the spatial dependence of TDOS and pair correlation function in this setting, which we may expect to be different as a corollary to this section.

\section{Interacting QH Edge Hamiltonian and Superconducting Proximity effect }\label{3}

\begin{figure}
    \centering
    \includegraphics[scale=0.05]{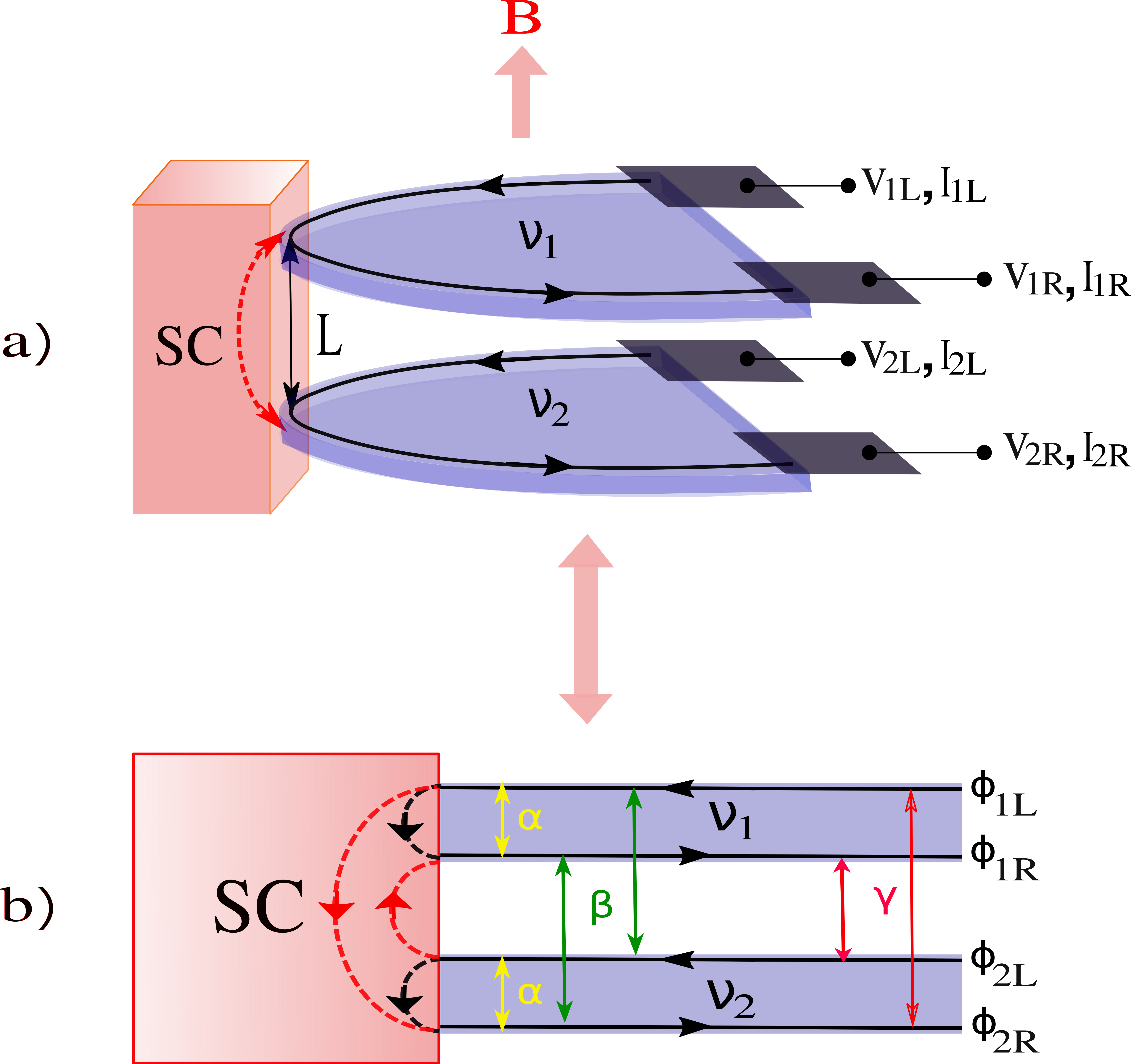}
    \caption{The schematic figure a) shows two QH states, with filling fraction $\nu_{1}$ and $\nu_{2}$ in a bilayer stacking induced by a uniform magnetic field $B$, strongly coupled to a superconductor (SC) at the apex. $V_{iR/L}$ and $I_{iR/L}$ are the voltages and the corresponding electronic current (satisfying the Hall relation) belonging to the right/left moving edge  of the $i^{th}$ QH state. The distance between the two QH layers at the superconducting junction is denoted by “$L$'' and is assumed to be less than the superconducting coherence length in order to facilitate the cross Andreev tunneling (denoted by dashed red line) between the edges of the two QH systems at the junction. In fig. b), $\phi_{iR/L}$ denotes the chiral bosonic field corresponding to the right/left moving field of the $i^{th}$ QH system. $\alpha,\beta,\gamma$ indicates the density-density interaction, where 1) $\alpha$ is the interaction between $\rho_{iR/L}$ and $\rho_{iL/R}$, 2) $\beta$ is the interaction between $\rho_{1R/L}$ and $\rho_{2R/L}$, and 3) $\gamma$ is the interaction between $\rho_{1R/L}$ and $\rho_{2L/R}$. The parameters, $\alpha,\beta,\gamma$ will symbolize the interaction strength in this article unless otherwise mentioned.  The dashed black line at the junction denotes the direct Andreev Reflection. The dashed red line at the junction denotes the cross Andreev tunneling. Here, the subscript “L" and “R" stand for
    the left moving fields flowing into the junction and the right moving fields flowing out of the junction.}
    \label{fig:setup}
\end{figure}

The model which is considered here is a similar to the one  mentioned in Ref.~\cite{Amulya2021} but with boundary condition corresponding to a junction of edge states of two QH system strongly coupled to a superconductor as shown in fig.~\ref{fig:setup}. We allow for repulsive density-density interaction to exist between all the QH edge states corresponding to the two QH layers. The chiral fermionic fields $\psi_{i R/L}$ can be expressed in terms of chiral bosonic fields $\phi_{iR/L}$ as $\psi_{i R/L} \sim F_{i R/L} e^{(\iota \phi_{i R/L}/\nu_{i})}$~\cite{Haldane_1981,Tgiamarchi2004,wen1990chiral,wen1990PRL,wen1991edge,wen1991gapless,wen1992theory,kane1992transport,von1998bosonization,Srao2002}, where the subscript R(L) describes right (left) moving fields and $\nu_{i}$ is the filling fraction of the $i^{th}$ QH layer. $F_{i R/L}$ is the Klein factor for right/left moving fields. The bosonized interacting QH edge Hamiltonian is given by

\begin{equation}
H = \frac{ \hbar v_{F} }{4 \pi} \sum_{ij=1}^{4} \int_{0}^{\infty} dx\:\frac{K_{ij}}{\sqrt{\nu_{i}\nu_{j}}} \:\;\partial_{x}\phi_{i}(x)\partial_{x}\phi_{j}(x)
\label{Eq:interacting Hamiltonian}
\end{equation}

where $v_{F}$ is the Fermi velocity and $ (\phi_{1}, \phi_{2}, \phi_{3}, \phi_{4}) = (\phi_{1R}, \phi_{2R}, \phi_{1L}, \phi_{2L})$. The matrix $K$ is given by 

\begin{equation}
K = \begin{pmatrix}
              1 & \beta & -\alpha & -\gamma \\
              \beta & 1 & -\gamma & -\alpha \\
              -\alpha & -\gamma & 1 & \beta \\
              -\gamma & -\alpha & \beta & 1
              \end{pmatrix}
\end{equation}

where

1)$\alpha$ is the interaction between the counter-propagating edge states in the same QH layer (intralayer interaction).

2)$\beta$ is the interaction between the co-propagating edge states of the different QH system (interlayer interaction).

3)$\gamma$ is the interaction between the counter-propagating edge states of the different QH system (interlayer interaction).

The commutation relation between the bosonic fields is given by $\left[ \phi_{iR/L}(x),\phi_{jR/L}(y) \right] = \pm i\pi \nu_{i}\delta_{ij}\mathrm{Sgn}(x-y)$. The electronic charge density operator is given by $\rho_{i R/L} = \pm (1/2\pi) \partial_{x}\phi_{iR/L}$. The interacting Hamiltonian can be diagonalized as done in the  ref~\cite{Amulya2021,das2009effect}, by expressing interacting bosonic fields in terms of Bg fields as 
\begin{equation}
\begin{pmatrix}
\bar{\phi}_{R} \\
\bar{\phi}_{L}
\end{pmatrix}_{(x,t)} = \begin{pmatrix}
                X_{1} & X_{2} \\
                X_{3} & X_{4}
                \end{pmatrix} \begin{pmatrix}
                              \tilde{\phi}_{R} \\
                              \tilde{\phi}_{L}
                              \end{pmatrix}_{(x,t)}~.
\label{Eq: couplet bar_tilde fields rlation}
\end{equation}

where, $\bar{\phi}_{R/L} = \left( \phi_{1 R/L}/\sqrt{\nu_{1}},\phi_{2 R/L}/\sqrt{\nu_{2}}\right)^{T}$. The commutation relation for Bg fields is then given by $\left[ \tilde{\phi}_{iR/L}(x),\tilde{\phi}_{jR/L}(y) \right]=\pm i\pi\delta_{ij}\mathrm{Sgn}(x-y)$. Superconducting boundary condition at the junction, is expressed as the current splitting matrix S  and corresponds to the different fixed points of the theory. Taking into account the fact that all the fields are defined from $x=0$ to $x=\infty$, the interacting bosonic fields $\phi_{iR/L}$ can be expressed in terms of left moving Bg field $\tilde{\phi}_{iL}$ as
\begin{eqnarray}
\phi_{R}(x,t) &=& M\left[T_{1} \tilde{\phi}_{L}(-x,t) + T_{2}\tilde{\phi}_{L}(x,t)\right], \nonumber\\
\phi_{L}(x,t) &=& M\left[T_{3} \tilde{\phi}_{L}(-x,t) + T_{4}\tilde{\phi}_{L}(x,t)\right],
\label{modified_bosonic_field} 
\end{eqnarray}
where, $\tilde{\phi}_{R/L}(x,t) = \left( \phi_{1R/L},\phi_{2R/L} \right)_{(x,t)}^{T}$, $\left[M\right]_{ij} = \sqrt{\nu_{i}}\delta_{ij}$ and
\begin{eqnarray}
T_{1} &=& X_{1}\left(X_{1} - \bar{S}X_{3}\right)^{-1}\left(\bar{S}X_{4}-X_{2}\right), \nonumber\\
T_{2} &=& X_{2}, \nonumber\\
T_{3} &=& X_{3}\left(X_{1} - \bar{S}X_{3}\right)^{-1}\left(\bar{S}X_{4}-X_{2}\right), \nonumber\\
T_{4} &=& X_{4},
\end{eqnarray}
where, matrix $\bar{S} = M^{-1}SM$. For a junction of edge states corresponding to two QH layers in the vicinity of a superconductor, it should be noted that there are only two fixed point corresponding to the superconducting boundary condition other than the conventional charge conserving fixed point discussed in Ref.~\cite{Amulya2021,das2006interedge,wen1990chiral,das2009effect,wen1990PRL,wen1991edge,wen1991gapless,wen1992theory,wen1994impurity,sen2008line,chklovskii1998consequences,sandler1999noise}. The two superconducting fixed points are given by the current splitting matrices as 
\begin{equation}
\mathrm{A_{2}}=\begin{pmatrix}
      -1 & 0 \\
      0 & -1 
      \end{pmatrix}
\label{Eq:S1 FP}
\end{equation}
\begin{equation}
\mathrm{CA_{2}} = \frac{-1}{\nu_{1}+\nu_{2}}\begin{pmatrix}
                                   \nu_{1} - \nu_{2} & 2\nu_{1} \\
                                   2\nu_{2} & \nu_{2} - \nu_{1}
                                   \end{pmatrix}
\label{Eq:S2 FP}
\end{equation}

The disconnected Andreev fixed point, $\mathrm{A_{2}}$, corresponds to the case when an incident electron current along a left moving edge of a QH layer perfectly reflects back as a hole current in the right moving edge of the same QH layer. The cross Andreev $\mathrm{CA_{2}}$ fixed point corresponds to the case when an incident electronic current along a left moving edge of a first QH layer with $\nu_{1} >\nu_{2}$, gets partially transmitted and partially reflected as a hole current at the junction. Also, an incident electronic current along the left moving edge of the second QH layer, with $\nu_{2}<\nu_{1}$, gets transmitted as hole current with larger amplitude than that of incident electron current and partially reflects back as an electronic current. For both the fixed points, the total current conservation at the junction is violated by the factor of $2$, such that, $\sum_{i=1}^{2}\;j_{i,L}(0) - j_{i,R}(0) = 2\sum_{i=1}^{2}\;j_{i,L}(0)$~\cite{Affleck2006junctions,Duality}. Taking into account the boundary condition, one can find the structure of the $\mathrm{CA_{2}}$ fixed point using the bosonic commutation relations as shown in Ref.~\cite{wen1990chiral,wen1990PRL,wen1991edge,wen1991gapless,wen1992theory,wen1994impurity,sen2008line,das_rao_sen_2006}.  

\subsection{Power-Law Dependence of TDOS}

The electronic tunneling density of states~\cite{agarwal2009enhancement, kane1992transport,Amulya2021} (\tdos) at energy $E$ and distance $x$ from the junction, is given by

\begin{eqnarray}
\rho(x,E) &=& \int^{\infty}_{-\infty} \langle 0\vert \psi(x,t)\psi^{\dagger}(x,0)\vert0\rangle e^{-iEt} dt.
\label{Eq:TDOS_Standard}
\end{eqnarray}

TDOS is calculated in the right moving edge, as they carry the information about the fixed point that the junction is tuned to. Using the bosonization formula, one can express Eq.~\ref{Eq:TDOS_Standard} in terms of bosonic fields as

\begin{equation}
\rho_{{i}}(x,E) \sim \int_{-\infty}^{\infty} dt \langle 0\vert e^{i\frac{\phi_{iR}(x,t)}{\nu_{i}}}e^{-i\frac{\phi_{iR}(x,0)}{\nu_{i}}}\vert0 \rangle e^{-iEt}, \nonumber
\end{equation}  

where $i$ is the index of the QH layer. The energy power law of the \tdos at the junction is denoted by $\Delta^{0}$. In the zero-energy limit, TDOS is enhanced when $\Delta^{0}-1<0$, is marginal when $\Delta^{0} = 1$ and is suppressed when $\Delta^{0}-1>0$. Here, we mainly focused on \tdos at the junction and its relative evolution at finite distance $x$ away from the junction. Before we go further, we first discuss the TDOS in the limit $x\rightarrow \infty$. The energy power law of the TDOS in this limit is denoted by $\Delta^{\infty}$, such that 

\begin{eqnarray}
\Delta_{i}^{\infty} =&& \frac{1}{2\nu_{i}} \left( \frac{1-\beta}{\sqrt{(1-\beta)^{2} - (\alpha - \gamma)^{2}}} \right. \nonumber\\
 && + \left. \frac{1+\beta}{\sqrt{(1+\beta)^{2} - (\alpha + \gamma)^{2}}} \right). 
\end{eqnarray}

$\Delta^{\infty}_{i}$ does not depend on the type of fixed point (superconducting or normal), as in the limit $x\rightarrow \infty$, $\langle \psi^{\dagger}_{iR}(x,t)\psi_{iR}(x,0) \rangle$ becomes insensitive to the boundary condition. Although, the bulk power law gets modified in presence of non-local interaction, it is still always greater than 1, i.e, $\Delta^{\infty} >1$, as a result of which TDOS in the limit $x\rightarrow \infty$ is always suppressed. We get the standard $1/\nu$  power-law suppression in TDOS for an edge of a fractional quantum Hall state in the limit $\alpha=\beta=\gamma = 0$~\cite{chang2003chiral} and also, in the case when $\alpha = \gamma = 0$ while  $\beta \neq 0$. The latter is due to the fact that the $\beta$ interaction corresponds to a forward scattering interaction and hence can result only in the renormalization of Fermi velocity but can not influence the power law of correlation functions.

The TDOS at finite distance $x$ from the junction has a power law dependence on $x$ in the limit $x << \mathrm{max}\lbrace \tilde{v}_{1},\tilde{v}_{2} \rbrace /E$, where $\tilde{v}_{i}$ is the renormalized velocity. We define a term, the `relative TDOS', as the ratio of the TDOS at finite distance $x$ and the TDOS at the junction  in the zero energy limit, i.e, $(\rho(x,E\rightarrow0)/\rho(x\rightarrow0,E\rightarrow 0))$. The relative TDOS has a pure spatial power law dependence in the $x<< \mathrm{max}\lbrace \tilde{v}_{1},\tilde{v}_{2} \rbrace /E$ limit and is given by
\begin{equation}
\frac{\rho_{i}(x,E\rightarrow 0)}{\rho_{i}(x=0,E\rightarrow 0)} \propto \left( \frac{\delta^{2} + 4x^{2}}{\delta^{2}}\right)^{ \varsigma_{i}}
\label{Eq:TDOS_x}
\end{equation} 
where $\varsigma_{i} = \sum_{j=1}^{2}\frac{[T1]_{ij}[T_{2}]_{ij}}{\nu_{i}}$ (see Appendix.~\ref{Appendix TDOS}). If the spatial power law of relative TDOS, $\varsigma_{i}$, is negative, then the relative TDOS decays as a power law of $x$ and is indicative of the fact that TDOS at the junction is less suppressed as compared to the TDOS in the $x\rightarrow\infty$ limit. On the other hand, if $\varsigma_{i}$ is positive, then the relative TDOS increases as a power law of $x$ and implies that the TDOS at the junction is more suppressed as compared to the TDOS in the $x\rightarrow\infty$ limit.

\subsection{ Scaling dimensions of perturbation operators at the junction } 
The stability of the superconducting junction fixed points, in the presence of non-local interaction, is determined by analyzing the scaling dimensions of all the physically relevant perturbations which can be switched on at the junction. In general, the stability at the junction is determined against the direct Andreev reflection (AR) operator, cross Andreev reflection (CAR) operator, intralayer quasi-particle backscattering operator and interlayer electron tunneling operator. The scaling dimensions of the operators can be calculated using Eq.~\ref{modified_bosonic_field} as follows

\begin{enumerate}
\item[(1)]
 The direct Andreev reflection (AR) operator, $\psi_{e,i R}(0)\psi_{e,iL}(0)$, has a  scaling dimension given by $(1/2)\sum_{k=1}^{2}(\Lambda_{AR, i}^{k})^{2}$, where
\begin{equation}
\Lambda_{AR,i}^{k} = \frac{1}{\sqrt{\nu_{i}}} (T_{1} + T_{2} + T_{3} + T_{4})_{ik}
\label{Eq: dAR}
\end{equation}

\item[(2)]
 The cross Andreev reflection (CAR) operator, $\psi_{e,i R}(0)\psi_{e,jL}(0)$, has a  scaling dimension given by $(1/2)\sum_{k=1}^{2}(\Lambda_{CAR, ij}^{k})^{2}$, where
\begin{equation}
\Lambda_{CAR,ij}^{k} = \frac{1}{\sqrt{\nu_{i}}} (T_{1} + T_{2})_{ik} + \frac{1}{\sqrt{\nu_{j}}}(T_{3} + T_{4})_{jk}
\label{Eq: dcar}
\end{equation}

\item[(3)]
 The intralayer quasiparticle backscattering operator, $\psi^{qp\dagger}_{i R}(0)\psi^{qp}_{i L}(0)$, has a  scaling dimension given by  $(1/2)\sum_{k=1}^{2}(\Lambda_{B, i}^{k})^{2}$, where
\begin{equation}
\Lambda_{B,i}^{k} = \sqrt{\nu_{i}}(T_{3} + T_{4} - T_{1} - T_{2})_{ik},
\label{Eq: dii}
\end{equation}

\item[(4)] 
The interlayer electron tunneling operator, $\psi^{\dagger}_{e,iR}(0)\psi_{e,j L}(0)$, has a scaling dimension given by $(1/2)\sum_{k=1}^{2}(\Lambda_{T, ji}^{k})^{2}$, where
\begin{equation}
\Lambda_{T,ij}^{k} = \frac{1}{\sqrt{\nu_{i}}} (T_{3} + T_{4})_{ik} - \frac{1}{\sqrt{\nu_{j}}} (T_{1} + T_{2})_{jk}. 
\label{Eq: dij}
\end{equation}
\end{enumerate}  

The junction fixed point is stable when the perturbation operators in the vicinity of the junction becomes irrelevant, that is, the corresponding scaling dimension of the all the physically relevant operators at the junction becomes more than unity simultaneously. In the vicinity of the junction $x=0$, for disconnected Andreev fixed point, $\mathrm{A_{2}}$, physically relevant perturbation operators are the CAR, electron tunneling and quasi-particle backscattering operators. For the cross Andreev fixed point, $\mathrm{CA_{2}}$, physically relevant  perturbation operators are the AR, electron tunneling, quasi-particle backscattering operators.

\subsection{Power Law Dependence of Induced Pair Amplitude}
In this section, we will evaluate the induced pair amplitude when the junction of chiral edges state corresponding to the two QH system are in proximity to a superconductor. In the subsequent sections, the aim is to study the effect of non-local interactions on the pair correlation function. We will compare the evolution of TDOS and the pair correlation function at finite distance $x$ from the junction. As a corollary to section~\ref{section 1}, we expect the spatial dependence of both, the TDOS and pair correlation function, to be different. It would be interesting to see if the presence of non-local interaction can reverse the order of the decay of TDOS and pair correlation function, as opposed to what was observed in section~\ref{section 1}. The pair amplitude is given by the anomalous pair correlation function, defined as $F_{ij}(x,t) = \langle \psi_{iR}(x,t^{+}) \psi_{jL}(x,t)\rangle$\footnote{Such pairing correlation functions have already been studied, both experimentally and theoretically, in Ref.\cite{Yacoby2020induced,michelsen2020current}}. There are two types of pair correlation functions, which are given by:

\begin{enumerate}
\item[(1)]
Local pair correlation function given by  $F_{ii}(x) = \langle\psi_{iR}(x,t^{+})\psi_{iL}(x,t)\rangle \sim \langle e^{ i \frac{\phi_{iR}(x,t^{+})}{\nu_{i}}} e^{ i \frac{\phi_{iL}(x,t)}{\nu_{i}}}\rangle$ for  $i \in \lbrace 1,2 \rbrace$. $F_{ii}(x)$ is calculated for the case when we have disconnected Andreev $\mathrm{A_{2}}$ fixed point as the boundary condition at the junction. 

\item[(2)]
Non-local pair correlation function given by $F_{12}(x) = \langle \psi_{iR}(x,t^{+})\psi_{jL}(x,t)\rangle \sim \langle e^{ i \frac{\phi_{iO}(x,t^{+})}{\nu_{i}}} e^{ i \frac{\phi_{jI}(x,t)}{\nu_{j}}}\rangle$ for $i \neq j$ and  $\lbrace i,j \rbrace \in \lbrace 1,2 \rbrace$. $F_{12}(x)$ is calculated when we have cross Andreev fixed point as the boundary condition at the junction.
\end{enumerate}

The pair correlation function is calculated between the right moving edge of the $i^{th}$ QH system and the left moving edge of the $j^{th}$ QH system, in the limit $T\rightarrow 0$ and $L\rightarrow \infty$ and is given by
\begin{eqnarray}
F_{ij}(x) &=& \left(\frac{1}{2\pi\delta}\right) \Pi_{k=1}^{2}\left(-\frac{i 2 \pi}{L}\right)^{\Lambda^{0}_{ijk}}\left(i \delta\right)^{\Lambda^{1}_{ijk}} \nonumber \\
&& \times \left(2x + i\delta\right)^{\Lambda^{2}_{ijk}}\left(-2x + i\delta\right)^{\Lambda^{3}_{ijk}}
\label{Eq:Fij_x_general}
\end{eqnarray}
where,
\begin{eqnarray}
\Lambda^{0}_{ijk} &=& \frac{1}{2}\left[ \frac{[T_{1}]_{ik} + [T_{2}]_{ik}}{\sqrt{\nu_{i}}} + \frac{[T_{3}]_{jk} + [T_{4}]_{jk}}{\sqrt{\nu_{j}}}\right]^{2} \nonumber \\
\Lambda^{1}_{ijk} &=& \frac{1}{2}\left[ \left( \frac{[T_{1}]_{ik}}{\sqrt{\nu_{i}}} + \frac{[T_{3}]_{jk}}{\sqrt{\nu_{j}}}\right)^{2} + \left( \frac{[T_{2}]_{ik}}{\sqrt{\nu_{i}}} + \frac{[T_{4}]_{jk}}{\sqrt{\nu_{j}}}\right)^{2}  \right]\nonumber \\
\Lambda^{2}_{ijk} &=& \frac{1}{2}\left[ \frac{[T_{1}]_{ik}  [T_{2}]_{ik}}{\nu_{i}} + \frac{[T_{3}]_{jk}  [T_{4}]_{jk}}{\nu_{j}} + \frac{2[T_{1}]_{ik}[T_{4}]_{jk}}{\sqrt{\nu_{i}\nu_{j}}}\right] \nonumber\\
\Lambda^{3}_{ijk} &=& \frac{1}{2}\left[ \frac{[T_{1}]_{ik}  [T_{2}]_{ik}}{\nu_{i}} + \frac{[T_{3}]_{jk}  [T_{4}]_{jk}}{\nu_{j}} + \frac{2[T_{2}]_{ik}[T_{3}]_{jk}}{\sqrt{\nu_{i}\nu_{j}}}\right]. \nonumber\\
\label{Eq: Pair_Amp_power_law}
\end{eqnarray}

In general, the short-wavelength cut-off $\delta$ is taken to be the order of superconducting coherence length $\xi_{SC} = \hbar v_{F}/\Delta_{SC}$~\cite{fazio1997properties}, where $\Delta_{SC}$ is the superconducting gap. In the limit $\Delta_{SC} \rightarrow \infty$, $\delta \rightarrow 0$, and can be considered as the lowest length scale available in the system. The induced pair amplitude is given by the real part of $F_{ij}(x)$, which, in the limit $x>>\delta$, has a spatial power law dependence given by (see Appendix~\ref{Appendix PA}). Here, we define a term, the `relative pair amplitude', as the ratio of the real part of the pair correlation function at finite distance $x$ to the real part of the pair correlation function at the junction $x=0$, i.e, $\mathrm{Re\left[F_{ij}(x)\right]/Re\left[F_{ij}(x\rightarrow 0)\right]}$. The relative pair correlation function has a pure power law dependence in the $x>>\delta$ limit and is given by
\begin{equation}
\frac{\mathrm{Re}\left[ F_{ij}(x)\right]}{\mathrm{Re}\left[F_{ij}(x\rightarrow 0 )\right]} \propto \left(\frac{\delta^{2} + 4x^{2}}{\delta^{2}}\right)^{\Lambda_{ij}} 
\label{Eq: Pair_Amp_x}
\end{equation}

where, $\Lambda_{ij} = \sum_{k=1}^{2} (\Lambda^{2}_{ijk} + \Lambda^{3}_{ijk})/2$. It can be seen from Eq.~\ref{Eq:TDOS_x}, Eq.~\ref{Eq: Pair_Amp_power_law}
and Eq.~\ref{Eq: Pair_Amp_x} that the spatial power law of relative TDOS and relative pair correlation function are very different in their composition owing to the fact that the Green's function for both, the TDOS and $F_{ij}(x)$ are different. Hence, the evolution of both, the TDOS and $F_{ij}(x)$ is distinct at finite distance $x$ away from the junction. The power laws are fixed point specific and hence will be analyzed in detail in the subsequent section for different scenarios.

\section{Superconducting junction of QH Edge states with equal filling fraction $\nu_{1} = \nu_{2}$}\label{4}

We will first present a brief review of Ref.~\cite{Amulya2021} for the case when $\nu_{1} = \nu_{2} \in \lbrace 1, 1/3\rbrace$ in the presence of non-local interaction with current conserving boundary condition as the fixed point and contrast this with the case when we have superconducting boundary condition as the fixed point. To summarize, for a fixed point corresponding to the current conserving boundary condition in the presence of non-local interactions, we can have simultaneous TDOS enhancement at the junction and stability of the fixed point, provided the symmetry between the two QH layers about the junction is broken. In the case when $\nu_{1} = \nu_{2}$, this is achieved by having asymmetry in the interaction between the counter-propagating edges of the same QH layer. Then the entanglement due to non-local interaction between the two QH layers can stabilize the junction in the interaction parameter regime, where we have enhancement in TDOS at the junction in the zero energy limit.

\begin{figure*}[t]
\centering
\includegraphics[scale = 0.04]{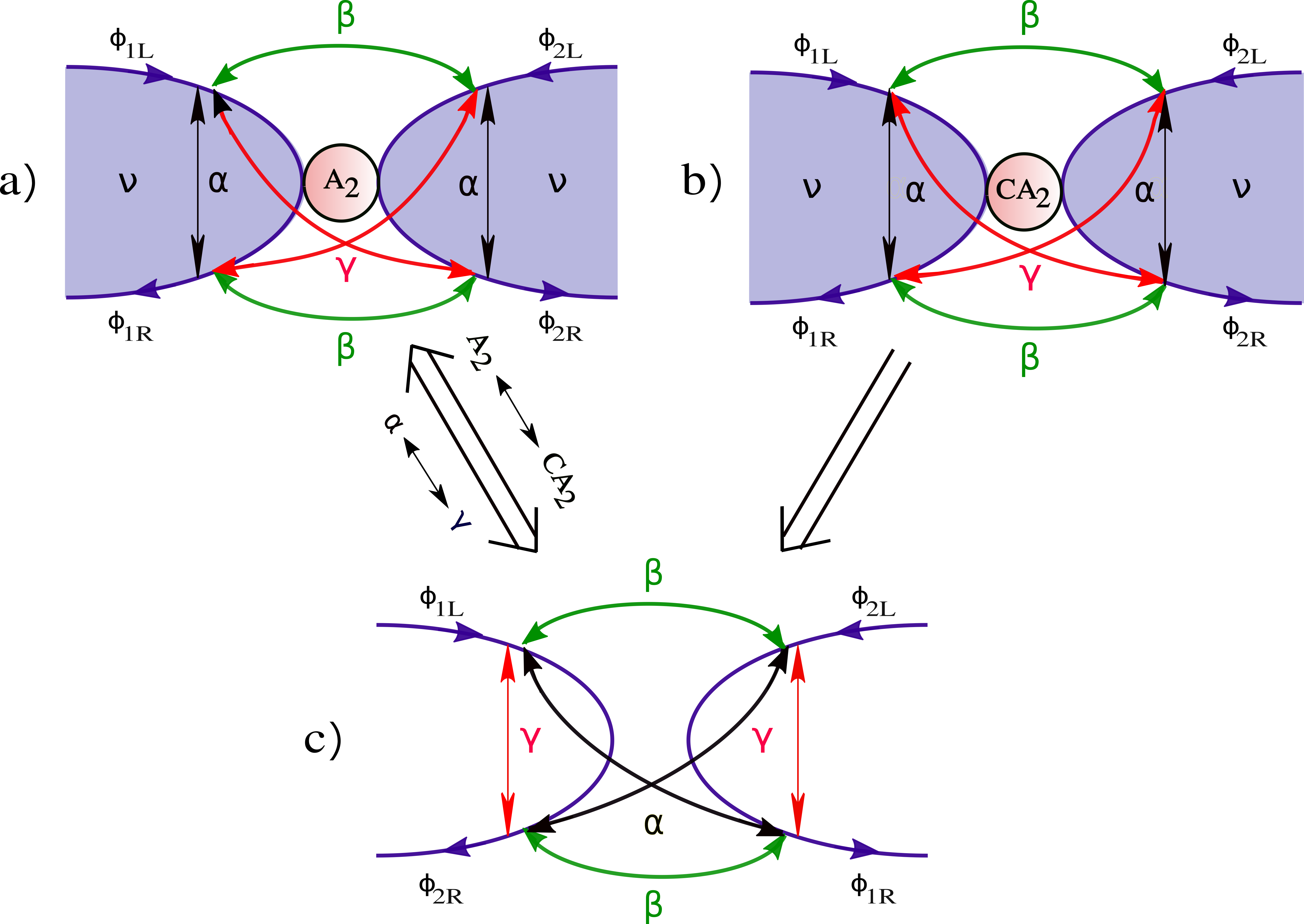}
\caption{The schematic figure a) and b) shows the unfolded version of the junction of SC with the edge states of two FQH system with equal filling fraction $\nu$, tuned to $\mathrm{A_{2}}$ and $\mathrm{CA_{2}}$ fixed point respectively. $\alpha,\beta,\gamma$ interactions are shown between the edges states about the junction. The $\mathrm{A_{2}}$ fixed point is given by the BC, $\phi_{i,R}(0) = -\phi_{i,L}(0)$ at the junction for $i\in \lbrace 1,2 \rbrace$. The $\mathrm{CA_{2}}$ fixed point is given by the BC, $\phi_{1/2,R}(0) = -\phi_{2/1,L}(0)$ at the junction. Figure c) shows a version of $\mathrm{CA_{2}}$ fixed point (figure b)) where the edge $\phi_{2R}$ is folded on the side of $\phi_{1L}$ and $\phi_{1R}$ is folded on the side of $\phi_{2L}$ to create an equivalent setup as shown in figure a). Comparison between figure a) and figure c) shows the symmetry between $\mathrm{A_{2}}$ and $\mathrm{CA_{2}}$ fixed point in the $\alpha \leftrightarrow \gamma$ exchange.}    
\label{fig:nu_symm_A2_CA2}
\end{figure*}

Coming back to the case when the edge states of the two QH system are strongly coupled to a superconductor, there are two possible current non-conserving fixed points, which in the case of $\nu_{1} = \nu_{2}$, are given by

\begin{eqnarray}
\mathrm{A_{2}} = \begin{pmatrix}
		-1 & 0 \\
		0 & -1	
		\end{pmatrix};\;\;\; \mathrm{CA_{2}} = \begin{pmatrix}
									0 & -1 \\
									-1 & 0
								   \end{pmatrix}.
\end{eqnarray}

For $\mathrm{A_{2}}$ fixed point, the TDOS energy power law exponent at the junction is denoted by $\Delta^{0}_{A_{2}}$ and is the same for both the right moving edges corresponding to the two QH layers. $\Delta^{0}_{A_{2}}$ is given by

\begin{eqnarray}
\Delta_{A_{2}}^{0} =&& \frac{1}{2\nu} \left( \sqrt{\frac{1 - \alpha -\beta +\gamma}{1+\alpha - \beta - \gamma}}
+ \sqrt{\frac{1-\alpha+\beta - \gamma}{1+\alpha + \beta +\gamma}} \right) \nonumber\\.
\label{Eq: Delta_A2_symm}
\end{eqnarray}

\begin{figure*}[t]
\centering
\includegraphics[scale = 0.35]{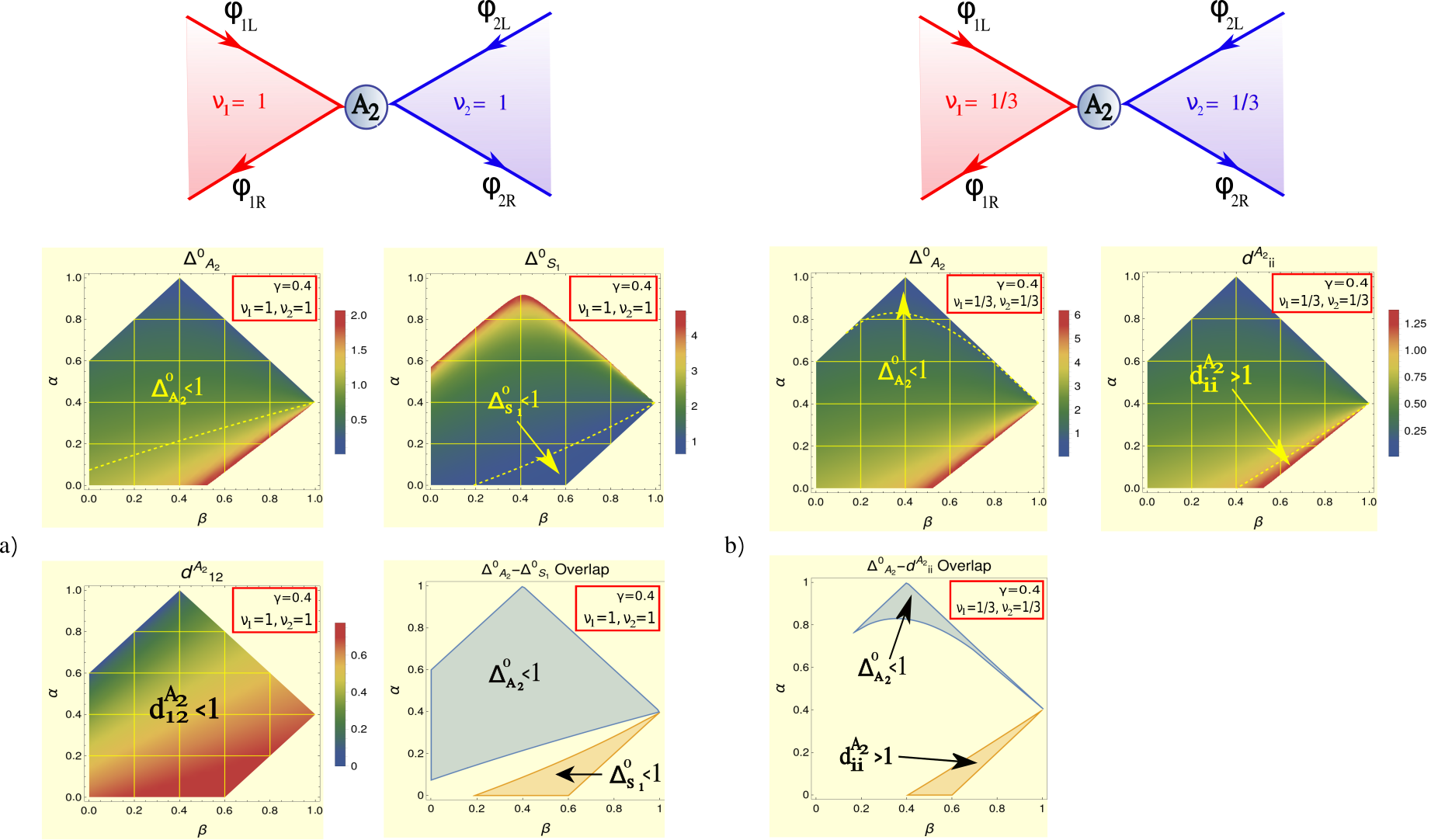}
\caption{The Schematic picture on the top shows the unfolded version of the direct Andreev reflection  $\mathrm{A_{2}}$ fixed point for a junction of two QH system in bi-layer stacking. Figure (a) shows a junction of $\nu_{1} =\nu_{2} = 1$ QH system in presence of interactions symmetric in the layer. The four density plots correspond to $\Delta^{0}_{A_{2}}$, $\Delta^{0}_{S_{1}}$, $d^{A_{2}}_{12}$ and the plot for the interaction parameter region for which the TDOS is enhanced in right moving  $\nu=1$ edge when the junction is tuned to $A_{2}$ and $S_{1}$ fixed points. Density plots are plotted for $\gamma = 0.4$. Figure (b) shows the junction of $\nu_{1} = \nu_{2} = 1/3$ QH system in presence of interactions symmetric in the layer. The three density plots correspond to $\Delta^{0}_{A_{2}}$, $d^{A_{2}}_{ii}$, and the interaction parameter region for which TDOS is enhanced, and the junction fixed point is stable against quasi-particle backscattering operator for $\gamma = 0.4$.}
\label{fig:A2_v_FP_plots}
\end{figure*}

\begin{figure*}[t]
\centering
\includegraphics[scale = 0.3]{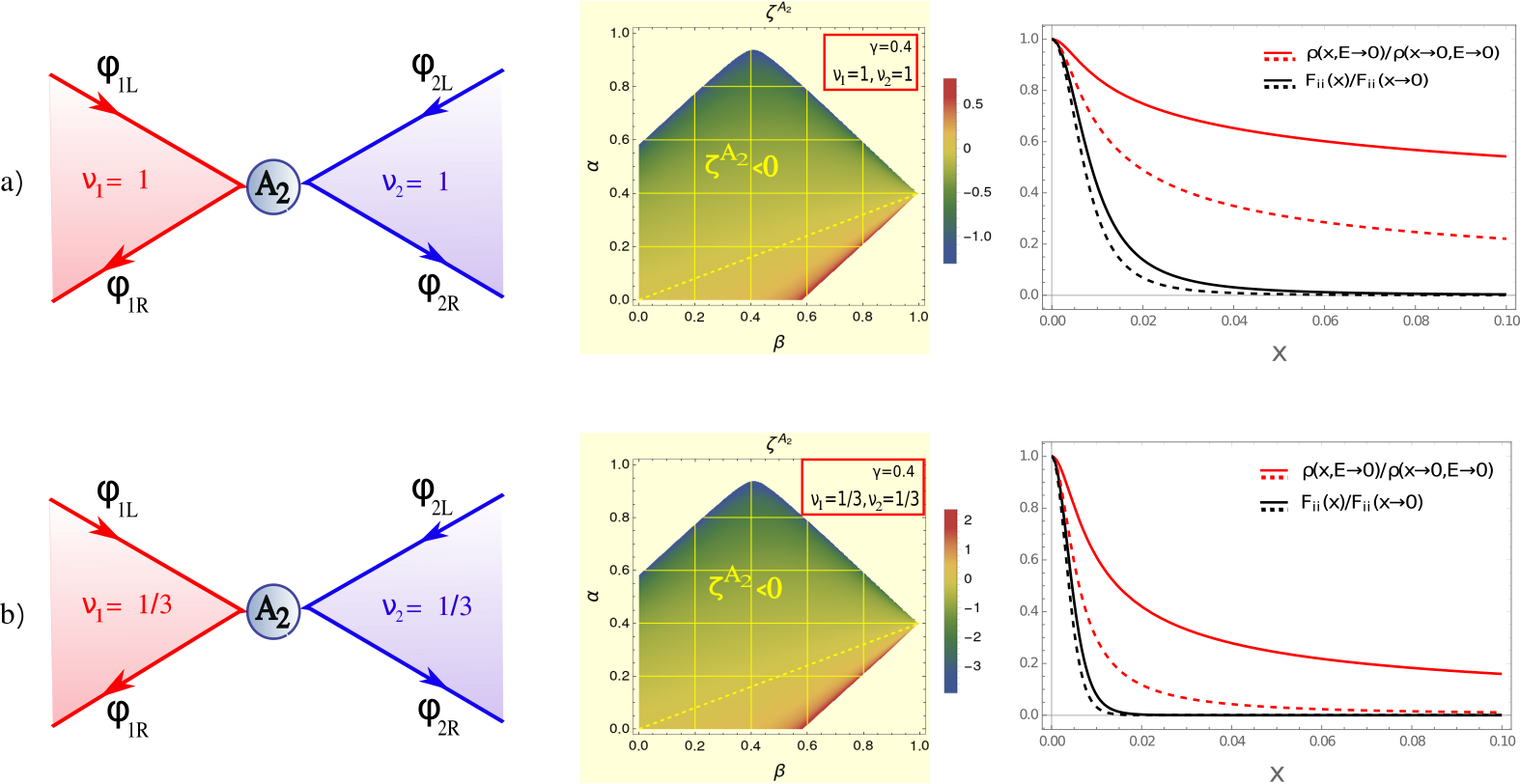}
\caption{The Schematic picture on the left shows the unfolded version of the direct Andreev reflection $\mathrm{A_{2}}$ fixed point for a junction of two QH system in bi-layer stacking. Figure (a) and (b) shows a junction of $\nu_{1} =\nu_{2} = 1$ and $\nu_{1} =\nu_{2} = 1/3$ QH system, respectively, in presence of interactions symmetric in layer. Density plots corresponds to the spatial power law of relative TDOS for $\alpha = 0.4$. The Dashed line corresponds to $\varsigma^{A_{2}} = 0$ and the interaction parameter region for which $\varsigma^{A_{2}} < 0$ is shown. The rightmost plots show the decay profile of relative TDOS, $\rho(x,E \rightarrow 0)/\rho(x\rightarrow0,E\rightarrow0)$ (shown in red), and relative pair correlation function $F_{ii}(x)/F_{ii}(x\rightarrow0)$ (shown in black). The solid line shows relative TDOS (red line) and relative pair correlation function (black line) decay for $\alpha = 0.2,\beta=\gamma=0$. The Dashed line shows relative TDOS (red) and relative pair correlation function (black) decay for $\alpha=0.5,\beta=\gamma=0.4$.}
\label{fig:TDOS_PA_A2_nu1_eq_nu2_plots}
\end{figure*}

\begin{figure*}[t]
\centering
\includegraphics[scale = 0.35]{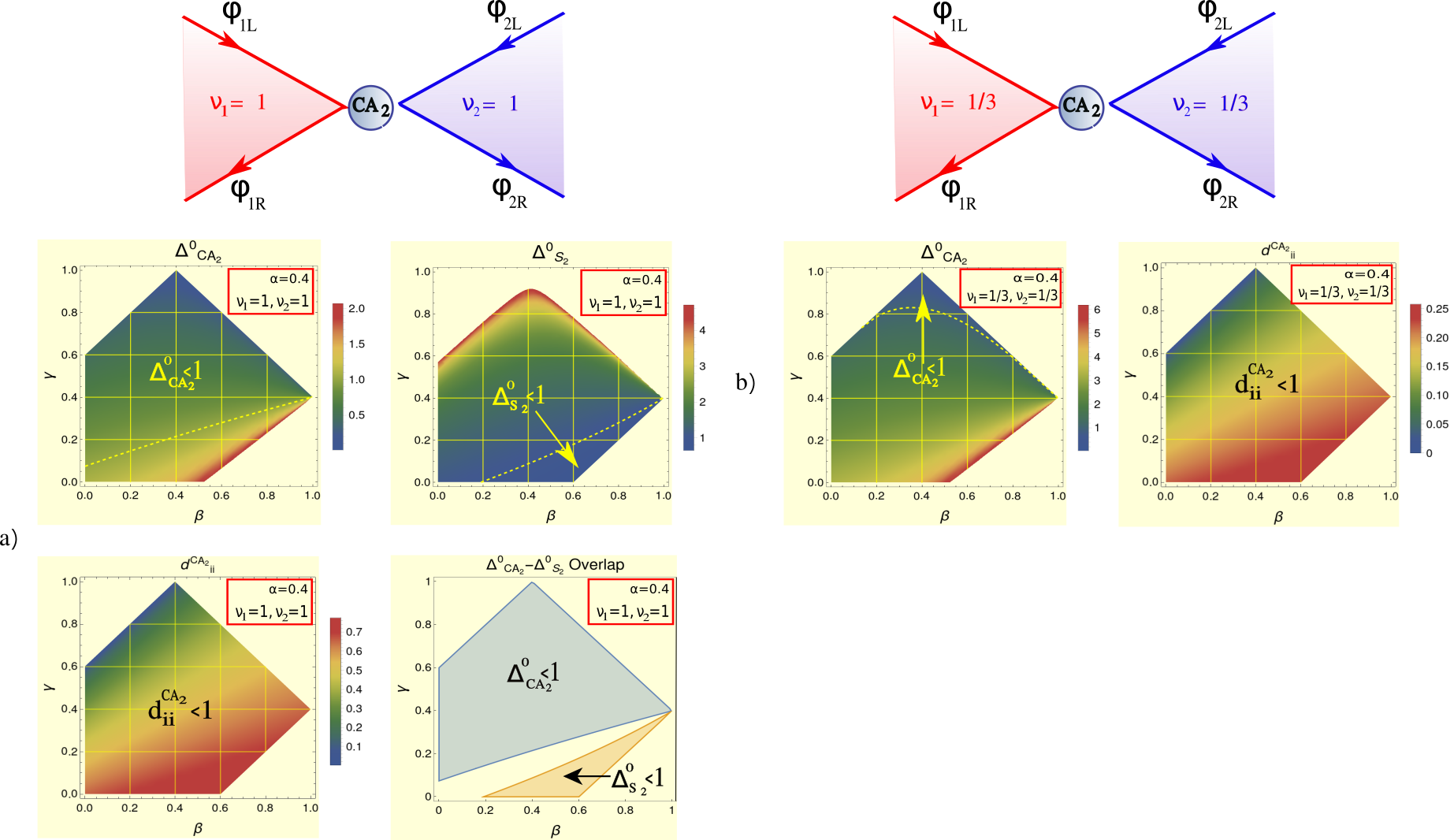}
\caption{The Schematic picture on the top shows the unfolded version of the Cross Andreev reflection  $\mathrm{CA_{2}}$ fixed point for a junction of two QH system in bi-layer stacking. Figure (a) shows a junction of $\nu_{1} =\nu_{2} = 1$ QH system in presence of interactions symmetric in layer. The four density plots correspond to $\Delta^{0}_{CA_{2}}$, $\Delta^{0}_{S_{2}}$, $d^{CA_{2}}_{ii}$ and the plot for the interaction parameter region for which the TDOS is enhanced in $\nu=1$ edge when the junction is tuned to $\mathrm{CA_{2}}$ and $S_{2}$ fixed points. Density plots are plotted for $\alpha = 0.4$. Figure (b) shows the junction of $\nu_{1} = \nu_{2} = 1/3$ QH system in presence of interactions symmetric in layer. The two density plots correspond to $\Delta^{0}_{CA_{2}}$, $d^{CA_{2}}_{ii}$ for $\alpha = 0.4$.}
\label{fig:CA2_v_FP_plots}
\end{figure*}

\begin{figure*}[t]
\centering
\includegraphics[scale = 0.3]{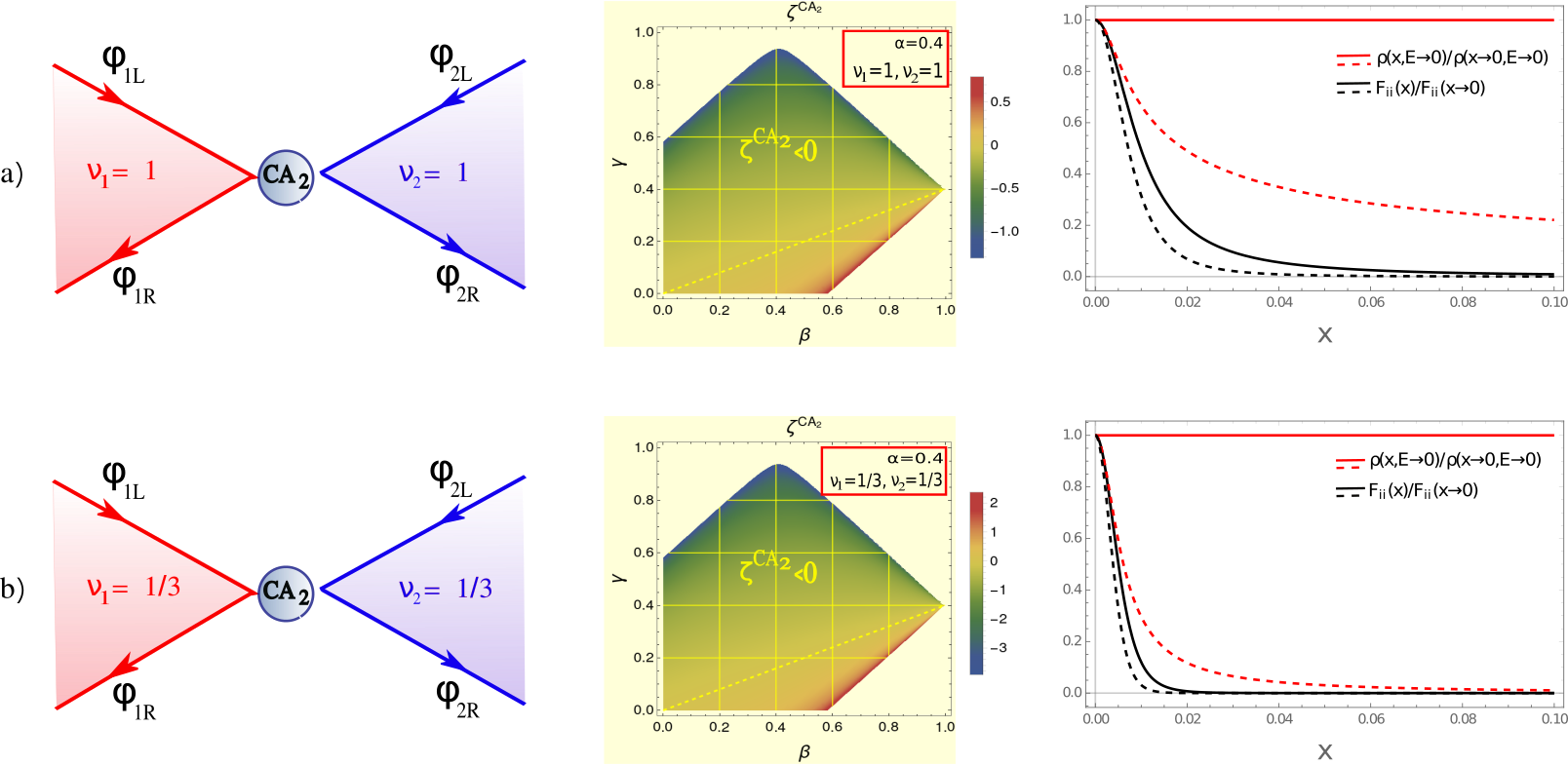}
\caption{The Schematic picture on the left shows the unfolded version of the Cross Andreev reflection $\mathrm{CA_{2}}$ fixed point for a junction of two QH system in bi-layer stacking. Figure (a) and (b) shows a junction of $\nu_{1} =\nu_{2} = 1$ and $\nu_{1} =\nu_{2} = 1/3$ QH system, respectively, in presence of interactions symmetric in layers. Density plots corresponds to the spatial power law of relative TDOS for $\alpha = 0.4$. The Dashed line corresponds to $\varsigma^{CA_{2}} = 0$ and the interaction parameter region for which $\varsigma^{CA_{2}} < 0$ is shown. The rightmost plots show the decay profile of relative TDOS, $\rho(x,E \rightarrow 0)/\rho(x\rightarrow0,E\rightarrow0)$ (shown in red), and relative pair correlation function $F_{12}(x)/F_{12}(x\rightarrow0)$ (shown in black). The solid line shows decay of relative TDOS (red line) and relative pair correlation function (black line) for $\alpha = 0.2,\beta=\gamma=0$. The Dashed line shows decay of relative TDOS (red) and relative pair correlation function (black) for $\alpha=\beta=0.4,\gamma=0.5$.}
\label{fig:TDOS_PA_CA2_nu1_eq_nu2_plots}
\end{figure*}

In the absence of non-local interactions, i.e, $\beta = \gamma =0$, $\Delta_{A_{2}}^{0} = g/\nu$ and the TDOS shows enhancement at the junction for $g < \nu$ in the zero energy limit. 

Comparing $\Delta^{0}_{A_{2}}$ with the TDOS power law exponent for the current conserving disconnected fixed point, $S_{1}$ (where $S_{1} = \mathbb{I}_{2\times 2}$), denoted by $\Delta^{0}_{S_{1}}$, we get the symmetry relation between the two fixed points counterparts in $\alpha \leftrightarrow -\alpha$ and $\gamma \leftrightarrow -\gamma$ exchange, such that,

\begin{equation}
\Delta^{0}_{A_{2}}(\alpha,\beta,\gamma)=\Delta^{0}_{S_{1}}(-\alpha,\beta,-\gamma)
\label{Eq: Duality_nu1_nu2_delta}
\end{equation} 

Eq.~\ref{Eq: Duality_nu1_nu2_delta} can be thought of as an extension of the symmetry relation as given in Eq.~\ref{Eq:symmetry_SC-LL}, to the case when we have the superconducting junction of fractional QH edge states with non-local interactions between them. This equation implies that the TDOS enhancement, which we get in the case of repulsive $\alpha,\gamma$ interaction ($\alpha,\gamma>0$) for $\mathrm{A_{2}}$ fixed point, is the same as that of the one we get for $S_{1}$ fixed point in presence of attractive $\alpha,\gamma$ interaction ($\alpha,\gamma < 0$). Note that the $\alpha,\gamma$ denotes the interaction between the counter-propagating edge modes of the system. Naively, it is expected that the $\beta$ interaction (interaction between the co-propagating modes) is not of relevance for Eq.~\ref{Eq: Duality_nu1_nu2_delta}, as the interaction between the co-propagating modes belongs to the pure forward scattering and should not directly influence the superconducting correlations.

As a result of the symmetry relation, the interaction parameter regime, in which TDOS shows enhancement for $\mathrm{A_{2}}$ and $\mathrm{S_{1}}$ fixed point, are mutually exclusive and separated by a set of intermediate interaction parameters for which TDOS is suppressed for both the fixed points (see fig.~\ref{fig:A2_v_FP_plots} a). This separation in parameter space is a direct consequence of non-local interaction present in the system. In the absence of non-local interactions, the TDOS enhanced region in parameter space for $\mathrm{A_{2}}$ and $S_{1}$ fixed point are adjacent to each other and are identified by $g<1$ and $g>1$, respectively. 

In order to understand the interplay of various interaction  parameters which may lead to the enhancement of TDOS, we study the $\Delta^{0}_{A_{2}}$ in the weak $\alpha,\gamma$ limit by carrying out an expansion of $\Delta_{A_{2}}^{0}(\alpha,\beta,\gamma)$ around $(\alpha=0,\beta,\gamma=0)$ to the leading orders in $\alpha$ and $\gamma$, such that $\Delta^{0}_{A_{2}}$ in the weak $\alpha,\gamma$ limit is given by
\begin{equation}
\Delta_{A_{2}}^{0} \simeq \frac{1}{\nu_{i}}\left( 1+\frac{\beta\gamma - \alpha}{1-\beta^{2}} \right) .
\label{pertur}
\end{equation}

It is interesting to note that the non-local interaction in this limit adversely affects the \tdos enhancement at the junction, while it had a favorable effect on the corresponding charge conserving $S_{1}$ fixed point, studied in Ref.~\cite{Amulya2021}. This can be understood as the direct consequence of the symmetry relation between the $\mathrm{A_{2}}$ and $S_{1}$ fixed points.

Consider the case of $\nu_{1}=\nu_{2}=1$ in the weak $\alpha,\gamma$ limit. In this limit, TDOS shows enhancement when $\alpha>\beta\gamma$. The minimal requirement for TDOS to show enhancement is the presence of only $\alpha$ interaction with $\beta , \gamma = 0$, as was reported in Ref.~\cite{winkelholz1996anomalous}. In the presence of finite $\alpha,\beta$ interaction with $\gamma = 0$, TDOS enhancement is further amplified (stronger power law divergence), beyond what was observed for local $\alpha$ interaction. Note that from Eq.~\ref{pertur}, an increase in $\beta$, starting from $\beta=0$ with $\gamma \neq 0$, can
lead to a crossover from enhancement to suppression in TDOS in the weak $\alpha,\gamma$ limit. 

In the non perturbative limit (for arbitrary values of $\alpha,\gamma$), TDOS shows enhancement in large $\alpha,\beta$ limit (see fig.~\ref{fig:A2_v_FP_plots} a) and b)). Even for the case of $\nu_{1}=\nu_{2}=1/3$, TDOS shows enhancement at the junction, although in the strong $\alpha,\beta$ limit (see fig.
~\ref{fig:A2_v_FP_plots}b)), which was impossible irrespective of the strength of $\alpha,\beta,\gamma$ in the case when we have a normal $S_{1}$ fixed point as the BC at the junction as shown in Ref.~\cite{Amulya2021}.

Next, we study the evolution of relative TDOS and the relative local pair correlation function for $\mathrm{A_{2}}$ fixed point as a function of distance $x$ from the junction. This can be done by comparing the spatial power laws as given in Eq.~\ref{Eq:TDOS_x} and Eq.~\ref{Eq: Pair_Amp_x}. The spatial power law for the relative TDOS and the relative pair correlation function is denoted by $\varsigma^{A_{2}}$ and $\Lambda^{A_{2}}{ii}$ respectively and are as follows

\begin{eqnarray}
\varsigma^{A_{2}} &=& \frac{1}{4\nu}\left[\frac{\gamma - \alpha}{\sqrt{(1-\beta)^{2} -(\alpha-\gamma)^{2} }}\right. \nonumber\\
&& \left.- \frac{\alpha + \gamma}{\sqrt{(1+\beta)^{2} - (\alpha + \gamma)^{2}}}\right] \nonumber\\
\Lambda^{A_{2}}_{ii} &=& -\frac{1}{4\nu}\left[\sqrt{\frac{1+\alpha-\beta-\gamma}{1-\alpha-\beta+\gamma}} + \sqrt{\frac{1+\alpha+\beta+\gamma}{1-\alpha+\beta-\gamma}}\right].\nonumber\\
\label{Eq:TDOS_PA_x_A2}
\end{eqnarray}

As can be seen from Eq.~\ref{Eq:TDOS_PA_x_A2}, the algebraic dependence of the two power laws on the interaction parameters are very different from each other. The relative local pair correlation function decreases as a power law at finite distance $x$ from the junction, as $\Lambda^{A_{2}}_{ii}$ is always negative irrespective of the type of the interaction (repulsive or attractive). On the other hand, the relative TDOS can show a transition from an increasing function to a decreasing function of $x$ about $\varsigma^{A_{2}}(\alpha,\beta,\gamma) = 0$ in the interaction parameter space. In the interaction parameter regime where TDOS is less suppressed at the junction with respect to the bulk, the induced pair amplitude function $F_{ii}(x)$ decays more rapidly as compared to TDOS at finite $x$ (see fig.~\ref{fig:TDOS_PA_A2_nu1_eq_nu2_plots}).

The stability of the $\mathrm{A_{2}}$ fixed point is determined against the cross Andreev reflection (CAR) operator $\psi_{1R}(0)\psi_{2L}(0)$, quasi-particle backscattering operator $\psi^{qp \dagger}_{iR}(0)\psi^{qp}_{iL}(0)$ and electron tunneling $\psi^{\dagger}_{1R}(0)\psi_{2L}(0)$ operator at the junction. The scaling dimension corresponding to these perturbation operators is given by $d^{A_{2}}_{CAR}$, $d^{A_{2}}_{ii}$ and $d^{A_{2}}_{12}$ respectively, and is given by,

\begin{eqnarray}
&& d^{A_{2}}_{CAR} = \frac{1}{\nu} \sqrt{\frac{1-\alpha-\beta +\gamma}{1+\alpha-\beta-\gamma}}, \nonumber\\
&& d^{A_{2}}_{ii} = \nu\left( \sqrt{\frac{1-\alpha -\beta + \gamma}{1+\alpha - \beta -\gamma}} + \sqrt{\frac{1-\alpha + \beta -\gamma}{1+\alpha + \beta +\gamma}} \right), \nonumber\\
&& d^{A_{2}}_{12} = \frac{1}{\nu} \sqrt{\frac{1-\alpha+\beta -\gamma}{1+\alpha+\beta+\gamma}}.
\end{eqnarray}

For $\mathrm{A_{2}}$ fixed point, with $\nu_{1}=\nu_{2}=1$, the stability at the junction gets compromised due to electron tunneling operator (see fig.~\ref{fig:A2_v_FP_plots}a). On the other hand, for $\nu_{1}=\nu_{2}=1/3$, the $\mathrm{A_{2}}$ fixed point becomes unstable against quasi-particle back-scattering operators in the interaction parameter regime where TDOS is enhanced at the junction (see fig.~\ref{fig:A2_v_FP_plots}b).

Importantly, for the case of $\nu_{1} = \nu_{2}$, we note that there exist a symmetry relation between the $\mathrm{A_{2}}$ fixed point and the $\mathrm{CA_{2}}$ fixed point in the $\alpha \leftrightarrow \gamma$ exchange. This implies that as we change from disconnected $\mathrm{A_{2}}$ to strongly coupled $\mathrm{CA_{2}}$ fixed point, the roles of $\alpha$ and $\gamma$ interactions gets interchanged (see fig.~\ref{fig:nu_symm_A2_CA2}). As a result of which, the energy power law exponent of TDOS, $\Delta^{0}_{CA_{2}}$, the spatial power law of the relative TDOS $\varsigma^{CA_{2}}$ and relative induced pair-amplitude $\Lambda^{CA_{2}}_{12}$, are given by

\begin{eqnarray}
\Delta^{0}_{CA_{2}}(\alpha,\beta,\gamma) &=& \Delta^{0}_{A_{2}}(\gamma,\beta,\alpha)\nonumber\\
\varsigma^{CA_{2}}(\alpha,\beta,\gamma) &=& \varsigma^{A_{2}}(\gamma,\beta,\alpha) \nonumber\\
\Lambda^{CA_{2}}_{12}(\alpha,\beta,\gamma) &=&
\Lambda^{A_{2}}_{11}(\gamma,\beta,\alpha)
\label{Eq:symm_A2_CA2_nu}
\end{eqnarray}

Note that the symmetry exist between the local induced pair amplitude for $\mathrm{A_{2}}$ fixed point and the non-local induced pair-amplitude for $CA_{2}$ fixed point. As can be seen from fig.~\ref{fig:TDOS_PA_A2_nu1_eq_nu2_plots} and fig.~\ref{fig:TDOS_PA_CA2_nu1_eq_nu2_plots}, the relative TDOS and relative induced pair amplitude follows the same behavior if the interaction parameters are changed from $\alpha = 0.5,\beta=0.4,\gamma=0.4$ for $\mathrm{A_{2}}$ fixed point to $\alpha = 0.4,\beta=0.4,\gamma=0.5$ for $\mathrm{CA_{2}}$ fixed point. Also, from Eq.~\ref{Eq: Delta_A2_symm} and Eq.~\ref{Eq:symm_A2_CA2_nu}, we note that in the weak $\alpha,\gamma$ interaction limit, for $\nu=1$, TDOS shows enhancement for $\gamma > \beta\alpha$. In general, the TDOS enhancement for $\mathrm{CA_{2}}$ fixed point is supported in the large $\gamma$ and small $\alpha$ limit (see fig.~\ref{fig:CA2_v_FP_plots}).

The stability of the $\mathrm{CA_{2}}$ fixed point is determined against direct Andreev reflection (AR) operator $\psi_{iR}\psi_{iL}$, electron tunneling operator $\psi^{ \dagger}_{1R}\psi_{2L}$ and quasiparticle backscattering operator $\psi^{qp \dagger}_{iR}\psi^{qp}_{iL}$ at the junction. The scaling dimension of these perturbation operators at the junction is given by $d^{CA_{2}}_{AR}$, $d^{CA_{2}}_{12}$ and $d^{CA_{2}}_{ii}$ respectively. We note that the scaling dimension of the backscattering (quasi-particle and Andreev reflection) operator and the tunneling (electron and cross Andreev reflection) operator for the two fixed points, $\mathrm{CA_{2}}$ and $\mathrm{A_{2}}$, are also related through a symmetry relation and are given by

\begin{eqnarray}
d^{CA_{2}}_{AR}(\alpha,\beta,\gamma) &=& d^{A_{2}}_{CAR}(\gamma,\beta,\alpha) \nonumber\\
d^{CA_{2}}_{12}(\alpha,\beta,\gamma) &=& \frac{1}{\nu^{2}}d^{A_{2}}_{ii}(\gamma,\beta,\alpha) \nonumber\\
d^{CA_{2}}_{ii}(\alpha,\beta,\gamma) &=& \nu^{2} d^{A_{2}}_{12}(\gamma,\beta,\alpha)
\end{eqnarray}

Note that the $\mathrm{CA_{2}}$ fixed point is unstable against quasi-particle backscattering operator for both, $\nu_{1}=\nu_{2}=1$ and $1/3$ (see fig.~\ref{fig:CA2_v_FP_plots}a and ~\ref{fig:CA2_v_FP_plots}b).

In the next section, we add further twist to the system by having different filling fraction in the two QH layers such that $\nu_{1} \neq \nu_{2}$ with $\nu_{1,2} \in \lbrace 1,1/3\rbrace$ and analyze the effect of unequal filling fraction on TDOS enhancement and stability of the superconducting junction fixed point in the presence of non-local interaction. 

\section{Superconducting junction of QH Edge states with unequal filling fraction $\nu_{1} \neq \nu_{2}$ }\label{5}

It was noted in Ref.~\cite{Amulya2021} that for the fixed points corresponding to the current conserving junction of edge states of two QH system with unequal filling fractions, $\nu_{1} \neq \nu_{2}$, one can have simultaneous TDOS enhancement and stability of the junction fixed point in presence of symmetric non-local interaction. In this section, we will analyze how the current non-conserving superconducting junction changes the scenario for TDOS and stability in the presence of non-local interaction for $\nu_{1}\neq \nu_{2}$. 

To be specific, we will focus on the junction of $\nu_{1} = 1$ and $\nu_{2} = 1/3$. The two Andreev fixed points corresponding to Eq.~\ref{Eq:S1 FP} and Eq.~\ref{Eq:S2 FP} are given by
\begin{equation}
A_{2} = \begin{pmatrix}
		-1 & 0 \\
		 0 & -1
		\end{pmatrix};\;\;\mathrm{CA_{2}} = \frac{-1}{2}\begin{pmatrix}
										 1 & 3 \\
										 1 & -1
										 \end{pmatrix}
\label{Eq: FP_nu1_nu2}										 
\end{equation}

\begin{figure*}[t]
\centering
\includegraphics[scale = 0.35]{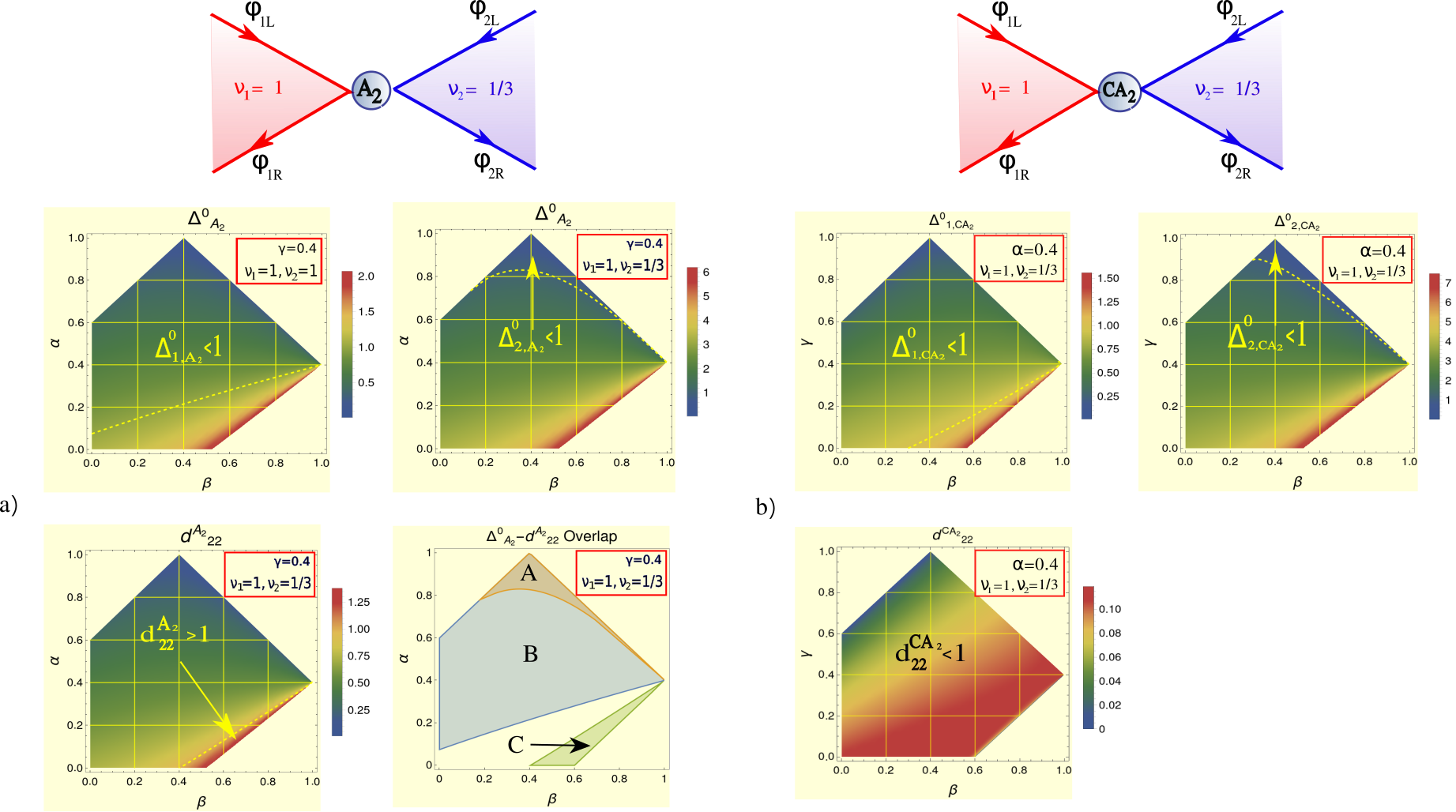}
\caption{The Schematic picture on the top shows the unfolded version of the junction of $\nu_{1} = 1,\nu_{2}=1/3$ QH system in a bilayer stacking tuned to $A_{2}$ and $\mathrm{CA_{2}}$ fixed point. Figure (a) shows the junction tuned to $\mathrm{A_{2}}$ fixed point. The four density plots corresponding to $\Delta^{0}_{1,A_{2}}$, $\Delta^{0}_{2,A_{2}}$, $d^{A_{2}}_{22}$ and the plot showing the interaction parameter region for which TDOS is enhanced and junction is stable with respect to quasi-particle backscattering operator in the $\nu_{2}=1/3$ QH layer for $\gamma = 0.4$. In the fourth plot, region A corresponds to $(\Delta^{0}_{1A_{2}} < 1,\Delta^{0}_{2,A_{2}} < 1)$, region B corresponds to $(\Delta^{0}_{1A_{2}} < 1,\Delta^{0}_{2,A_{2}} > 1)$ and region C corresponds to $d^{A_{2}}_{22}>1$. Figure (b) shows a junction of $\nu_{1} = 1$ and $\nu_{2} = 1/3$ QH system tuned to the $\mathrm{CA_{2}}$ fixed point. The three density plots corresponds to $\Delta^{0}_{1,CA_{2}}$, $\Delta^{0}_{2,CA_{2}}$ and $d^{CA_{2}}_{22}$ for $\alpha = 0.4$.}
\label{fig:nu1_neq_nu2_FP_plots}
\end{figure*}

\begin{figure*}[t]
\centering
\includegraphics[scale = 0.35]{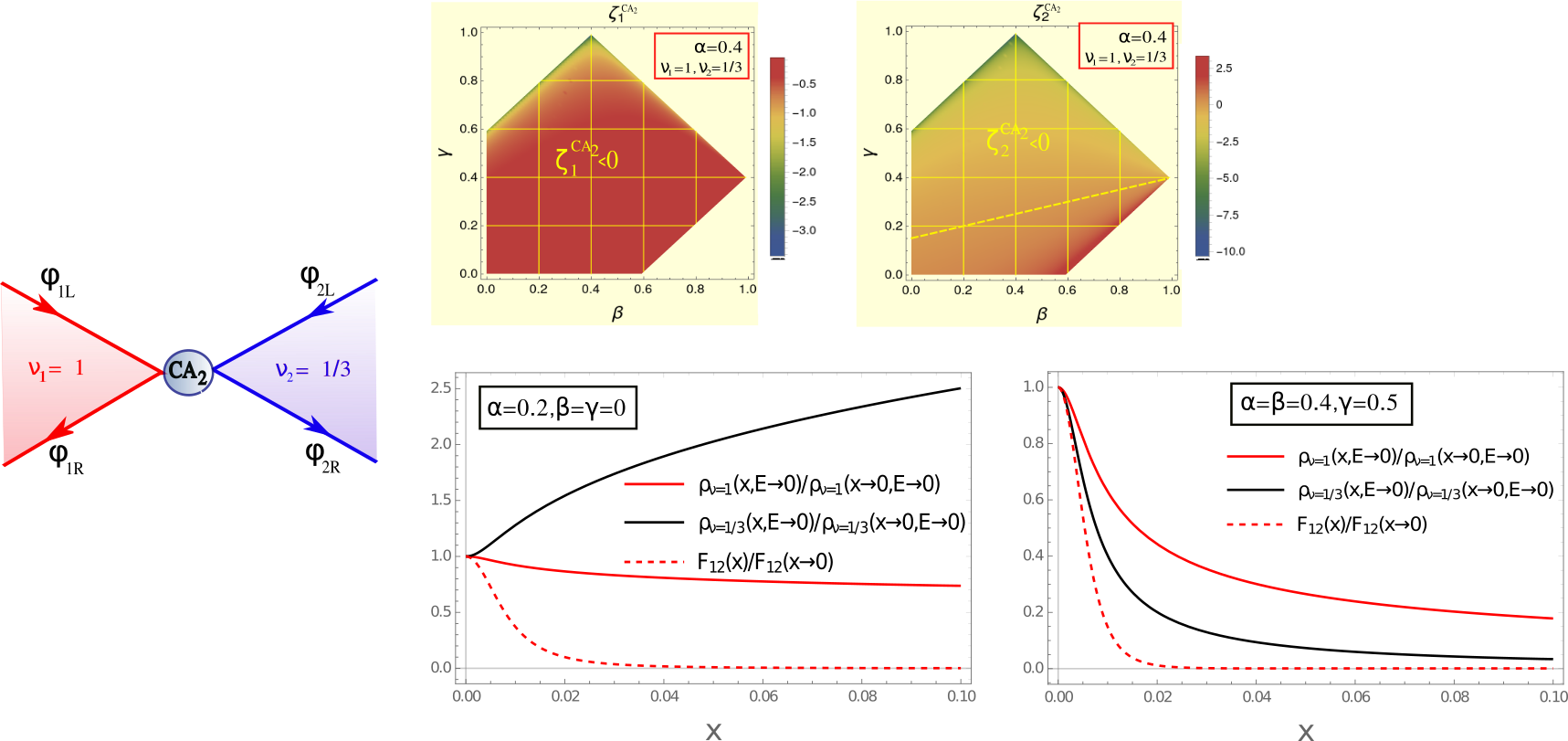}
\caption{The Schematic picture on the left shows the unfolded version of the Cross Andreev reflection $\mathrm{CA_{2}}$ fixed point for a junction of $\nu_{1}=1$ and $\nu_{2}=1/3$ QH system in bi-layer stacking. Density plots corresponds to the spatial power law of relative TDOS calculated for $\alpha = 0.4$. The Dashed line corresponds to $\varsigma^{CA_{2}}_{2} = 0$ and the interaction parameter region for which $\varsigma^{CA_{2}}_{i} < 0$ ($i\in \lbrace 1,2 \rbrace$) is shown. The two plots in the bottom shows the spatial profile of relative TDOS, $\rho(x,E \rightarrow 0)/\rho(x\rightarrow0,E\rightarrow0)$ (shown in solid red for $\nu_{1}=1$ edge and in solid  black for $\nu_{2}=1/3$ edge), and the relative pair correlation function $F_{12}(x)/F_{12}(x\rightarrow0)$ (shown in dashed red). The spatial profile plots for relative TDOS and relative pair correlation function are plotted for $\alpha = 0.2,\beta=\gamma=0$ and also for $\alpha=\beta=0.4,\gamma=0.5$.}
\label{fig:TDOS_PA_x_CA2_nu1_neq_nu2_plots}
\end{figure*}

Note that for $\mathrm{A_{2}}$ fixed point, the power law of any correlation function calculated between the modes of the same QH layer (for example, $\langle \psi^{\dagger}_{iR}(x,t)\psi_{iR}(x,0) \rangle$ or $\langle \psi_{iR}(x,t^{+})\psi_{iL}(x,t) \rangle$ for the $i^{th}$ QH layer) does not depend on the filling fraction of the other QH layer even though they coupled through non-local interactions. The energy power law for TDOS, that is $\Delta^{0}_{A_{2}}$, the spatial power law of relative TDOS and the relative local pair correlation function, remains the same as given in Eq.~\ref{Eq: Delta_A2_symm} and Eq.~\ref{Eq:TDOS_PA_x_A2}, respectively. Simultaneous TDOS enhancement can be observed in both the right moving edges of $\nu_{1}=1$ and $\nu_{2}=1/3$ QH layers, in the strong $\alpha$ limit (see fig.~\ref{fig:nu1_neq_nu2_FP_plots}a). The symmetry relation as given in Eq.~\ref{Eq: Duality_nu1_nu2_delta} remains valid here, between the disconnected normal $S_{1}$ fixed point and the $\mathrm{A_{2}}$ fixed point, even when $\nu_{1}\neq \nu_{2}$.

The stability of the $\mathrm{A_{2}}$ fixed point is determined against CAR operator, quasi-particle backscattering operator and electron tunneling operator. The notations of the scaling dimension remains the same as used in the section~\ref{4}. $d^{A_{2}}_{CAR}$, $d^{A_{2}}_{ii}$ and $d^{A_{2}}_{12}$, in the case when $\nu_{1} \neq \nu_{2}$, is given by  

\begin{eqnarray}
d^{A_{2}}_{CAR} &=& \frac{1}{4} \left[ \left( \sqrt{\frac{1-\alpha-\beta +\gamma}{1+\alpha-\beta-\gamma}}\right)\left(\frac{1}{\sqrt{\nu_{1}}} + \frac{1}{\sqrt{\nu_{2}}}\right)^{2}\right. \nonumber\\
&&+ \left. \sqrt{\frac{1-\alpha+\beta -\gamma}{1+\alpha+\beta+\gamma}}\left(\frac{1}{\sqrt{\nu_{1}}} - \frac{1}{\sqrt{\nu_{2}}}\right)^{2}\right] \nonumber\\
d^{A_{2}}_{ii} &=& \nu_{i}\left[ \sqrt{\frac{1-\alpha -\beta + \gamma}{1+\alpha - \beta -\gamma}} + \sqrt{\frac{1-\alpha + \beta -\gamma}{1+\alpha + \beta +\gamma}} \right] \nonumber\\
d^{A_{2}}_{12} &=& \frac{1}{4}\left[ \sqrt{\frac{1-\alpha -\beta +\gamma}{1+\alpha - \beta -\gamma}}\left( \frac{1}{\sqrt{\nu_{1}}} - \frac{1}{\sqrt{\nu_{2}}}\right)^{2} \right. \nonumber \\
&& \left. +  \sqrt{\frac{1-\alpha + \beta -\gamma}{1+\alpha + \beta +\gamma}}\left( \frac{1}{\sqrt{\nu_{1}}} + \frac{1}{\sqrt{\nu_{2}}} \right)^{2} \right] 
\end{eqnarray}

The $\mathrm{A_{2}}$ fixed point is unstable against the quasi-particle backscattering operator which can be switched on between the edge states of $\nu_{2}=1/3$ QH layer at the junction, in the region where TDOS is enhanced at the junction in either of the two right moving edges (see fig.~\ref{fig:nu1_neq_nu2_FP_plots}a). Hence, we do not have simultaneous TDOS enhancement at the junction and stability of the $\mathrm{A_{2}}$ fixed point. On the other hand, as was reported in Ref.~\cite{Amulya2021}, for disconnected normal fixed point, we can have simultaneous TDOS enhancement in $\nu_{1}=1$ QH edge and the stable fixed point against electron tunneling operator (the only physically relevant operator) for some set of interaction parameters.

Now we focus on the junction of edge states corresponding to $\nu_{1}=1$ and $\nu_{2}=1/3$ QH system, tuned to $\mathrm{CA_{2}}$ fixed point. The TDOS energy power law at the junction corresponding to the $i^{th}$ QH layer is given by $\Delta^{0}_{i,CA_{2}}$. The algebraic expression of $\Delta^{0}_{i,CA_{2}}$ is long and hence is not shown here. But we can analyze $\Delta^{0}_{i,CA_{2}}$ in the weak $\alpha,\gamma$ limit by expanding $\Delta^{0}_{i,CA_{2}}$ about $(\alpha = 0,\beta,\gamma =0)$. $\Delta^{0}_{i,CA_{2}}$ in the weak $\alpha,\gamma$ limit is given by

\begin{eqnarray}
\Delta^{0}_{i,CA_{2}} &\simeq & \frac{1}{\nu_{i}} \left[ 1 + \bar{a}_{i}\left(\frac{\alpha - \beta\gamma}{1-\beta^{2}}\right)
 - \bar{d}\left(\frac{\gamma - \beta\alpha}{1-\beta^{2}}\right) \right], \nonumber\\
\label{Eq: TDOS_CA2}
\end{eqnarray}

where `$\bar{a}_{i}$' is the diagonal element of the current splitting matrix corresponding to $\mathrm{CA_{2}}$ fixed point (see Eq.~\ref{Eq:S2 FP}), such that $\bar{a}_{1} = -(\nu_{1}-\nu_{2})/(\nu_{1}+\nu_{2})$ and $\bar{a}_{2} = -(\nu_{2}-\nu_{1})/(\nu_{1}+\nu_{2})$ and contributes anti-symmetrically to $\Delta^{0}_{i,CA_{2}}$ in the $\nu_{1} \leftrightarrow \nu_{2}$ exchange. $\bar{d}$ is given by $2\sqrt{\nu_{1}\nu_{2}}/(\nu_{1}+\nu_{2})$ and contributes symmetrically to the $\Delta^{0}_{i,CA_{2}}$ in the $\nu_{1}\leftrightarrow\nu_{2}$ exchange. In the absence of non-local interaction, that is, $\beta = \gamma = 0$, $\Delta^{0}_{i,CA_{2}}$ is given by $\Delta^{0}_{i,CA_{2}} = \frac{1}{2\nu_{i}}\left( g+\frac{1}{g}-\bar{a}_{i}\left(g-\frac{1}{g}\right)\right)$, which in the case of $\nu_{1} = \nu_{2}$, gives the usual bulk TDOS power-law, $\Delta_{i,CA_{2}}^{0} = \frac{1}{2\nu_{i}}\left( g+\frac{1}{g}\right)$ as expected from the discussion in section~\ref{section 1}. In the presence of non-local interaction, TDOS shows enhancement in the strong $\gamma$ and weak $\alpha$ limit in both the $\nu_{1}=1$ and $\nu_{2}=1/3$ edge (see fig.~\ref{fig:nu1_neq_nu2_FP_plots}b).

The stability of the $\mathrm{CA_{2}}$ fixed point is determined against direct Andreev reflection (AR) operator $\psi_{iR}(0)\psi_{iL}(0)$, electron tunneling operator $\psi^{\dagger}_{1R}(0)\psi_{2L}(0)$ and quasiparticle backscattering $\psi^{\dagger}_{qp,iR}(0)\psi_{qp,iL}(0)$ operator at the junction. The scaling dimensions of the above-mentioned perturbation operators are given by $d^{CA_{2}}_{i,AR}$, $d^{CA_{2}}_{12}$ and $d^{CA_{2}}_{ii}$ respectively. Here, the scaling dimensions and consequently the stability of the $\mathrm{CA_{2}}$ fixed point is studied numerically, as the exact expression for the scaling dimensions of the perturbation operator are too lengthy to report. We note that the $\mathrm{CA_{2}}$ fixed point for the junction of edge states belonging to $\nu_{1}=1$ and $\nu_{2}=1/3$ QH systems, is unstable against the quasi-particle backscattering operator which can be switched on at the junction between the edge states of $\nu_{2}=1/3$ QH layer irrespective of the strength of the interaction parameters (see fig.~\ref{fig:nu1_neq_nu2_FP_plots}b). 

The evolution of relative TDOS and the relative non-local pair correlation function, for the junction of $\nu_{1} \neq \nu_{2}$ QH system tuned to $\mathrm{CA_{2}}$ fixed point, is studied numerically. The spatial power law for the relative TDOS for the $i^{th}$ QH layer and the relative non-local pair correlation function is denoted by $\varsigma^{\mathrm{CA_{2}}}_{i}$ and $\Lambda^{\mathrm{CA_{2}}}_{12}$, respectively. We note that the relative TDOS for the right moving edge of $\nu_{1}=1$ QH layer always decreases as a power law at finite distance $x$ from the junction for all possible ($\alpha,\beta,\gamma$) in repulsive regime (see fig.~\ref{fig:TDOS_PA_x_CA2_nu1_neq_nu2_plots}). The relative TDOS in the edge of $\nu_{2} = 1/3$ can show a transition from increasing function to decreasing function of distance $x$ about $\varsigma^{CA_{2}}_{2} = 0$ in the interaction parameter space (see fig.~\ref{fig:TDOS_PA_x_CA2_nu1_neq_nu2_plots}). As observed previously, in the parameter regime where TDOS at the junction is less suppressed in one of the edge than the TDOS in the bulk of the same QH system, the relative nonlocal pair correlation function decays rapidly  as compared to the relative TDOS at finite distance $x$ away from the junction (see fig.~\ref{fig:TDOS_PA_x_CA2_nu1_neq_nu2_plots}).

\section{Discussion and Conclusion}\label{6}
It is well known that, for a LL QW strongly coupled to superconductor, TDOS shows enhancement in the zero energy limit in the vicinity of the junction~\cite{winkelholz1996anomalous}. Recently, it has been established that the presence of non-local interactions between the LLs can also give rise to enhancement in TDOS for fixed points corresponding to the current conserving boundary condition at the junction~\cite{Amulya2021}. In this paper, we assumed a general scenario of a junction of edge states corresponding to two fractional QH systems strongly coupled to a superconductor. Additionally, we have allowed for non-local density-density interactions to exists between the edge states of the two fractional QH systems. Here, we have done a comprehensive study of the possible scenarios for enhancement in TDOS and stability of the fixed point in the parameter space of the local and non-local (bulk) interactions.

The conclusions derived from the paper are as follows:
\begin{itemize}
\item[1)] TDOS can show enhancement in the vicinity of the superconducting junction in the presence of non-local interactions. Even the highly suppressed $\nu=1/3$ edge shows the TDOS enhancement, although in the strong interaction regime. This should be contrasted with the current conserving BC studied in Ref.~\cite{Amulya2021}, where TDOS enhancement was not possible $\nu=1/3$ QH edge irrespective of the strength of the interaction parameters as long as it is in repulsive regime. 

\item[2)] Simultaneous TDOS enhancement and stability of the junction fixed point is impossible for a superconducting junction against all physically relevant perturbations that can be switched on at the junction. If we only consider the Andreev type instabilities, then we can have simultaneous TDOS enhancement and stability of the fixed point at the junction (see fig.~\ref{fig: nu1_neq_nu2_andreev_stability} and fig.~\ref{fig: nu1_eq_nu2_A2_andreev_stability} in Appendix~\ref{Appendix AR_CAR_instability}).

\item[3)] A LL QW in proximity to a superconductor shows TDOS enhancement at the junction. In the weak $\alpha$ limit ($\alpha <<1$), upon introducing non-local interaction, specifically, increasing $\beta$ while keeping $\gamma$ at $\gamma = 0$, can further boost the existing enhancement in TDOS for finite $\alpha$.

\item[4)] There exist a symmetry relation between the fixed point corresponding to current conserving boundary condition and superconducting boundary condition. As a result of which, upon the introduction of superconducting correlations at the junction, the interaction parameter regime in which TDOS is enhanced at the junction for the fixed point corresponding to current conserving boundary condition, begins to show suppression in TDOS in the zero energy limit.

\item[5)]
In general, the TDOS does not follow spatial power law behavior at finite distance away from the junction. Hence, we can identify a quantity  "relative TDOS", defined as $\rho(x,E)/\rho(x\longrightarrow 0,E)$, which shows a pure spatial power law dependence in the $E\longrightarrow 0$ limit. We observed that, in the interaction parameter regime in which TDOS is less suppressed at the junction with respect to the bulk, the relative pair amplitude decays more rapidly as compared to the relative TDOS even in the presence of non-local interactions irrespective of the fixed point.  
\end{itemize}

The system which can naturally hosts the interlayer density-density interaction between the LLs is the quantum Hall systems in a bilayer stacking~\cite{li2018gate_blqh,fu2021gapless_blqh,shibata2009fractional_blqh,jacak2017unconventional_blqh,eisenstein2014exciton_blqh,finck2011exciton_blqh}, which, in general, has the interlayer distance $'d'$ of the order of $d \sim 30$ nm between the two $\mathrm{GaAs}$ quantum wells~\cite{eisenstein2014exciton_blqh,finck2011exciton_blqh}. This, in proximity to a superconductor, with superconducting coherence length $\xi_{SC}$ of the order of $\xi_{SC}\sim 38 - 230$ nm~\cite{kittel,ashcroft}, can host the model which supports both the non-local interaction and interlayer Andreev tunneling at the junction. The junction of superconductor with chiral (quantum Hall edge states) and non-chiral LL  have been of recent interest owing to the possibilities of hosting majorana/parafermion zero modes \cite{fazio1996dc_SC_LL_josephson,haim2014time_SC_LL_josepson,alicea2011non_SC_LL_josephson,fazio1995josephson_SC_LL_josephson,alicea2012universal_SC_LL_josephoson,affleck2013topological_SC_LL_josephson,zuo2014crossed,sahu2018inter_exp,hou2016crossed_exp,wan2015induced_exp,lee2017inducing_exp,clarke2014exotic_exp,guiducci2019toward_exp,barkeshli2016charge_exp,rickhaus2012quantum_exp,amet2016supercurrent_exp,hatefipour2021induced_exp,carrega2019investigation_exp,sadagashi2018cooper,sadagashi2019dominant}. Large cross Andreev reflection has already been observed experimentally for a fractional QH edge in proximity to a superconductor~\cite{lee2017inducing_exp,michelsen2020current,sadagashi2018cooper,sadagashi2019dominant}. Such experimental advancement suggest that the technology required for the proposed setup is not a farfetched one. 
      
  \begin{acknowledgments}
A.R. acknowledges University Grants Commission, India, for support in the form of a fellowship.
S.D. would  like to acknowledge the MATRICS grant (Grant No. MTR/ 2019/001 043) from the Science and Engineering Research Board (SERB) for funding.
\end{acknowledgments}

\bibliography{bibfile}

\appendix

\section{TDOS spatial and Energy power laws}\label{Appendix TDOS}
The Green's function for the TDOS for the right moving edge of the $i^{th}$ QH system, $\langle \psi^{\dagger}_{iR}(x,t)\psi_{iR}(x,0) \rangle$ in the presence of non-local interaction, is given by

\begin{eqnarray}
\langle \psi^{\dagger}_{iR}(x,t)\psi_{iR}(x,0) \rangle &=& \frac{1}{2\pi\delta} \prod_{j=1}^{2}\left( \frac{i\delta}{-\tilde{v}_{j}t + i\delta}\right)^{\Gamma_{ij}}  \nonumber\\
&& \times\left(\frac{(i\delta)^{2}-4x^{2}}{(i\delta-\tilde{v}_{j}t)^{2}-4x^{2}}\right)^{\zeta_{ij}}, \nonumber\\
\label{Eq:Gnrl_TDOS_correlator}
\end{eqnarray}

where $\tilde{v}_{i}$ are the renormalized velocities in the presence of symmetric non-local interactions. $\Gamma_{ij}$ and $\zeta_{ij}$ is given by $\frac{\left[T_{1}\right]^{2}_{ij} + \left[T_{2}\right]^{2}_{ij}}{\nu_{i}}$ and $\frac{\left[T_{1}\right]_{ij}\left[T_{2}\right]_{ij}}{\nu_{i}}$, respectively, where the expressions of T1 and T2 can be found in the main text. $\sum_{j=1}^{2}\Gamma_{ij}$ does not depend on the type of fixed point (superconducting or normal). On the other hand, $\sum_{j=1}^{2}\zeta_{ij}$ depends on the specifications of the boundary fixed point. The fixed point contribution in the Green's function comes from the $\left(\frac{(i\delta)^{2}-4x^{2}}{(i\delta-\tilde{v}_{j}t)^{2}-4x^{2}}\right)^{\zeta_{ij}}$, which in the limit $x\rightarrow \infty$ reduces to 1, as a result of which TDOS power law becomes insensitive to the boundary condition at the junction. In the limit $x\rightarrow 0$, Green's function picks up the contribution from the boundary condition and is given by $\prod_{j=1}^{2}\frac{1}{2\pi\delta}\left(\frac{i\delta}{i\delta-\tilde{v}_{j}t}\right)^{\Gamma_{ij}+2\zeta_{ij}}$. In the both the limit,$x\rightarrow 0$ and $x\rightarrow \infty$, the Green's function has the form of 
\begin{equation}
\langle \psi^{\dagger}_{iR}(x,t)\psi_{iR}(x,0) \rangle \propto \prod_{j=1}^{2}\frac{1}{2\pi\delta}\left(\frac{i\delta}{i\delta-\tilde{v}_{j}t}\right)^{\Delta_{ij}}
\end{equation}

From Eq.~\ref{Eq:TDOS_Standard}, the TDOS, in the limit $x\rightarrow 0$ and $x\rightarrow\infty$, has the form of
\begin{eqnarray}
\rho_{i} (x,E) &=& \frac{1}{2\pi\delta}\int_{-\infty}^{\infty}\prod_{j=1}^{2}\left(\frac{i\delta}{i\delta-\tilde{v}_{j}t}\right)^{\Delta_{ij}}e^{iEt}dt \nonumber\\
\end{eqnarray}

The above equation, in the $E\rightarrow 0$ limit, gives the energy power law divergence in TDOS of the form of $E^{\sum_{j=1}^{2}\Delta_{ij}-1}$, if $\sum_{j=1}^{2}\Delta_{ij} < 1$. The TDOS energy power law at the junction, hence, is given by $\Delta^{0} = \sum_{j=1}^{2}\Delta_{ij}$. The spatial power law of the TDOS at finite distance $x$ away from the junction can be calculated from Eq.~\ref{Eq:Gnrl_TDOS_correlator} and Eq.~\ref{Eq:TDOS_Standard}, such that,

\begin{eqnarray}
\rho_{i}(x,E) &=& \int_{-\infty}^{\infty}\frac{1}{2\pi\delta} \prod_{j=1}^{2}\left( \frac{i\delta}{-\tilde{v}_{j}t + i\delta}\right)^{\Gamma_{ij}}  \nonumber\\
&& \times\left(\frac{(i\delta)^{2}-4x^{2}}{(i\delta-\tilde{v}_{j}t)^{2}-4x^{2}}\right)^{\zeta_{ij}} e^{iEt}dt \nonumber\\
\end{eqnarray}
By taking constant terms in $t$, outside the integral, we get the form of the integral as
\begin{eqnarray}
\rho_{i}(x,E) &=& F(x)\int_{-\infty}^{\infty} \prod_{j=1}^{2}\left( \frac{1}{-\tilde{v}_{j}t + i\delta}\right)^{\Gamma_{ij}}  \nonumber\\
&& \times\left(\frac{1}{(i\delta-\tilde{v}_{j}t)^{2}-4x^{2}}\right)^{\zeta_{ij}} e^{iEt}dt \nonumber\\
\label{Eq:TDOS_finite_x_int}
\end{eqnarray}
where, the function $F(x)$ is given by $\frac{1}{2\pi\delta}(i\delta)^{\sum_{j=1}^{2}\Gamma_{ij}}\left((i\delta)^{2}-4x^{2}\right)^{\sum_{j=1}^{2}\zeta_{ij}}$. Now, the integral in Eq.~\ref{Eq:TDOS_finite_x_int}, can be written as
\begin{eqnarray}
I = \int_{-\infty}^{\infty} && \prod_{j=1}^{2}\left( \frac{1}{-\tilde{v}_{j}t + i\delta}\right)^{\Gamma_{ij}}  \times\left(\frac{1}{i\delta-\tilde{v}_{j}t-2x}\right)^{\zeta_{ij}}\nonumber\\
&& \times\left(\frac{1}{i\delta-\tilde{v}_{j}t+2x}\right)^{\zeta_{ij}} e^{iEt}dt
\label{Eq:TDOS_x_a}
\end{eqnarray}
A substitution of $Et = T$ can be made, such that the integral in Eq.~\ref{Eq:TDOS_x_a}, becomes

\begin{eqnarray}
I = C(E)\int_{-\infty}^{\infty} && \prod_{j=1}^{2} \left(\frac{1}{T-\frac{i \delta E}{\tilde{v}_{j}}}\right)^{\Gamma_{ij}} \left(\frac{1}{T-\frac{i \delta E}{\tilde{v}_{j}} +\frac{2xE}{\tilde{v}_{j}}}\right)^{\zeta_{ij}} \nonumber\\
&& \times \left(\frac{1}{T-\frac{i \delta E}{\tilde{v}_{j}} -\frac{2xE}{\tilde{v}_{j}}}\right)^{\zeta_{ij}} dT e^{iT}
\end{eqnarray}

where, $C(E)$ is given by 
\begin{equation}
C(E) = \frac{1}{E}\prod_{j=1}^{2}\left(\frac{-E}{\tilde{v}_{j}}\right)^{\Gamma_{ij}+2\zeta_{ij}} 
\end{equation}

In the limit $x << \mathrm{max}\lbrace\tilde{v}_{1},\tilde{v}_{2}\rbrace/E$, integral $I(x,E)$, will have the form of 
\begin{eqnarray}
I(E) \sim C(E)\int_{-\infty}^{\infty} dT \prod_{j=1}^{2}\left(\frac{1}{T-\frac{i\delta E}{\tilde{v}_{j}}}\right)^{\Gamma_{ij} + 2\zeta_{ij}}.
\end{eqnarray}

Hence, in the limit, $x << \mathrm{max}\lbrace \tilde{v}_{1},\tilde{v}_{2}\rbrace/E$, the relative TDOS has a pure spatial power law dependence, given by
\begin{equation}
\frac{\rho(x,E)}{\rho(x\rightarrow 0,E)} \sim \frac{F(x)I(E)}{F(x\rightarrow 0)I(E)} = \left(\frac{\delta^{2} + 4x^{2}}{\delta^{2}}\right)^{\sum_{j=1}^{2}\zeta_{ij}}    
\end{equation}

\section{Pair Amplitude spatial power law}\label{Appendix PA}
In general, a pair correlation function calculated between the right moving edge of the $i^{th}$ QH system and the left moving edge of the $j^{th}$ QH system, in the limit $T\rightarrow 0$ and $L\rightarrow \infty$, is given by
\begin{eqnarray}
F_{ij}(x) &=& \left(\frac{1}{2\pi\delta}\right) \prod_{k=1}^{2}\left(\frac{ 2 \pi}{L}\right)^{\Lambda^{0}_{ijk}}\left(\delta\right)^{\Lambda^{1}_{ijk}} \nonumber \\
&& \times \left(-i2x + \delta\right)^{\Lambda^{2}_{ijk}}\left(i2x + \delta\right)^{\Lambda^{3}_{ijk}}
\label{Eq:fij_x_appendix}
\end{eqnarray}
where,
\begin{eqnarray}
\Lambda^{0}_{ijk} &=& \frac{1}{2}\left[ \frac{[T_{1}]_{ik} + [T_{2}]_{ik}}{\sqrt{\nu_{i}}} + \frac{[T_{3}]_{jk} + [T_{4}]_{jk}}{\sqrt{\nu_{j}}}\right]^{2} \nonumber \\
\Lambda^{1}_{ijk} &=& \frac{1}{2}\left[ \left( \frac{[T_{1}]_{ik}}{\sqrt{\nu_{i}}} + \frac{[T_{3}]_{jk}}{\sqrt{\nu_{j}}}\right)^{2} + \left( \frac{[T_{2}]_{ik}}{\sqrt{\nu_{i}}} + \frac{[T_{4}]_{jk}}{\sqrt{\nu_{j}}}\right)^{2}  \right]\nonumber \\
\Lambda^{2}_{ijk} &=& \frac{1}{2}\left[ \frac{[T_{1}]_{ik}  [T_{2}]_{ik}}{\nu_{i}} + \frac{[T_{3}]_{jk}  [T_{4}]_{jk}}{\nu_{j}} + \frac{2[T_{1}]_{ik}[T_{4}]_{jk}}{\sqrt{\nu_{i}\nu_{j}}}\right] \nonumber\\
\Lambda^{3}_{ijk} &=& \frac{1}{2}\left[ \frac{[T_{1}]_{ik}  [T_{2}]_{ik}}{\nu_{i}} + \frac{[T_{3}]_{jk}  [T_{4}]_{jk}}{\nu_{j}} + \frac{2[T_{2}]_{ik}[T_{3}]_{jk}}{\sqrt{\nu_{i}\nu_{j}}}\right]. \nonumber\\
\end{eqnarray}

The induced pair amplitude is given by the real part of the pair correlation function as given in Eq.~\ref{Eq:fij_x_appendix}. The Eq.~\ref{Eq:fij_x_appendix} can be rewritten as
\begin{eqnarray}
F_{ij}(x) &=& \left(\frac{1}{2\pi\delta}\right) \prod_{k=1}^{2}\left(\frac{ 2 \pi}{L}\right)^{\Lambda^{0}_{ijk}}\left(\delta\right)^{\Lambda^{1}_{ijk}} \nonumber \\
&& \times \left(\delta^{2} + 4x^{2}\right)^{\frac{\Lambda^{2}_{ijk} + \Lambda^{3}_{ijk}}{2}}e^{i\left( \theta_{-}\Lambda^{2}_{ijk} + \theta_{+}\Lambda^{3}_{ijk}\right)} \nonumber\\
\end{eqnarray}
where, $\theta_{+} = -\theta_{-}= \tan^{-1}\left(\frac{2x}{\delta}\right)$. In the limit $x>>\delta$, $\theta_{\pm} = \pm\frac{\pi}{2}$. Hence, the induced pair amplitude, $\mathrm{Re}\left[F_{ij}(x)\right]$, is given by
\begin{eqnarray}
\mathrm{Re}\left[F_{ij}(x)\right] &\sim& \left(\frac{1}{2\pi\delta}\right) \prod_{k=1}^{2}\left(\frac{ 2 \pi}{L}\right)^{\Lambda^{0}_{ijk}}\left(\delta\right)^{\Lambda^{1}_{ijk}} \nonumber \\
&& \times \left(\delta^{2} + 4x^{2}\right)^{\frac{\Lambda^{2}_{ijk} + \Lambda^{3}_{ijk}}{2}}\cos{\frac{\pi}{2}\left(\Lambda^{3}_{ijk} - \Lambda^{2}_{ijk}\right)}\nonumber\\
\end{eqnarray}

Then, the relative pair correlation function is given by

\begin{equation}
\frac{\mathrm{Re}\left[F_{ij}(x)\right]}{\mathrm{Re}\left[F_{ij}(x\rightarrow 0 )\right]} \sim \left(\frac{\delta^{2} + 4 x^{2}}{\delta^{2}}\right)^{\sum_{k=1}^{2}\frac{\Lambda^{2}_{ijk} + \Lambda^{3}_{ijk}}{2}}.    
\end{equation}

\section{Stability analysis against  Andreev type perturbations }\label{Appendix AR_CAR_instability}
\begin{figure*}
    \centering
    \includegraphics[scale = 0.35]{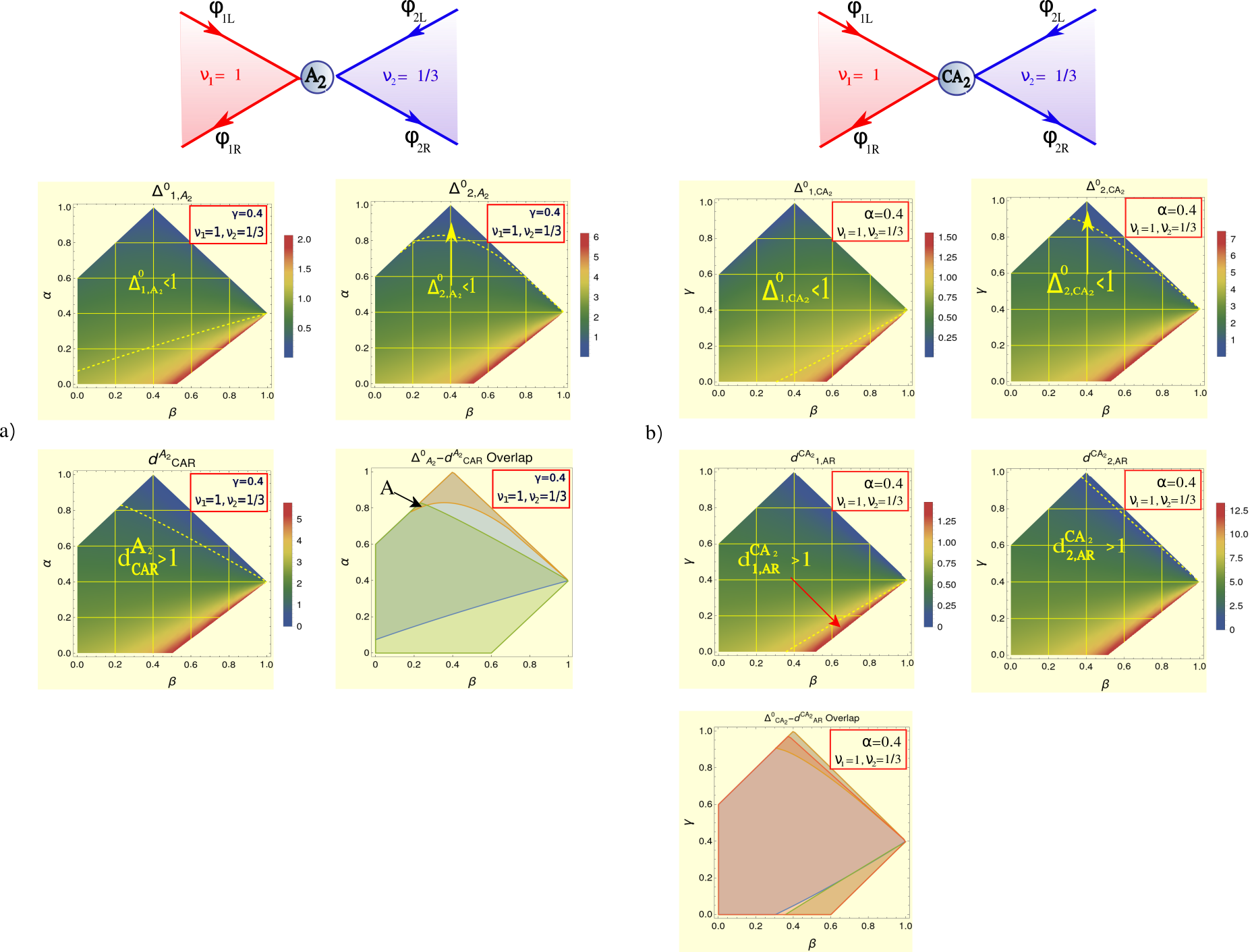}
    \caption{The schematic figure on the top shows the unfolded version of the junction of bilayer QH edge states with unequal filling fraction, $\nu_{1}=1$ and $\nu_{2}=1/3$, tuned to $\mathrm{A_{2}}$ and $\mathrm{CA_{2}}$ fixed points. In fig. a), the junction is tuned to $\mathrm{A_{2}}$ fixed point. The four plots correspond to $\Delta^{0}_{1,A_{2}}$, $\Delta^{0}_{2,A_{2}}$, $d^{A_{2}}_{CAR}$ and '$\Delta^{0}_{A_{2}}-d^{A_{2}}_{CAR}$ overlap' plot showing the interaction parameter region for which TDOS is enhanced and junction is stable against CAR operator for $\gamma = 0.4$. Region 'A' in $\Delta^{0}_{A_{2}}-d^{A_{2}}_{CAR}$ overlap plot  denotes the interaction parameters for which $\Delta^{0}_{1,A_{2}} < 1$, $\Delta^{0}_{2,A_{2}} < 1$, $d^{A_{2}}_{CAR} > 1$ simultaneously. In fig. b), the junction is tuned to $\mathrm{CA_{2}}$ fixed point. The five plots correspond to $\Delta^{0}_{1,CA_{2}}$, $\Delta^{0}_{2,CA_{2}}$, $d^{CA_{2}}_{1,AR}$, $d^{CA_{2}}_{2,AR}$ and '$\Delta^{0}_{CA_{2}}-d^{CA_{2}}_{AR}$ overlap' plot showing the interaction parameter region for which TDOS is enhanced and junction is stable against AR operator for $\alpha = 0.4$.     }
    \label{fig: nu1_neq_nu2_andreev_stability}
\end{figure*}

\begin{figure*}
    \centering
    \includegraphics[scale = 0.35]{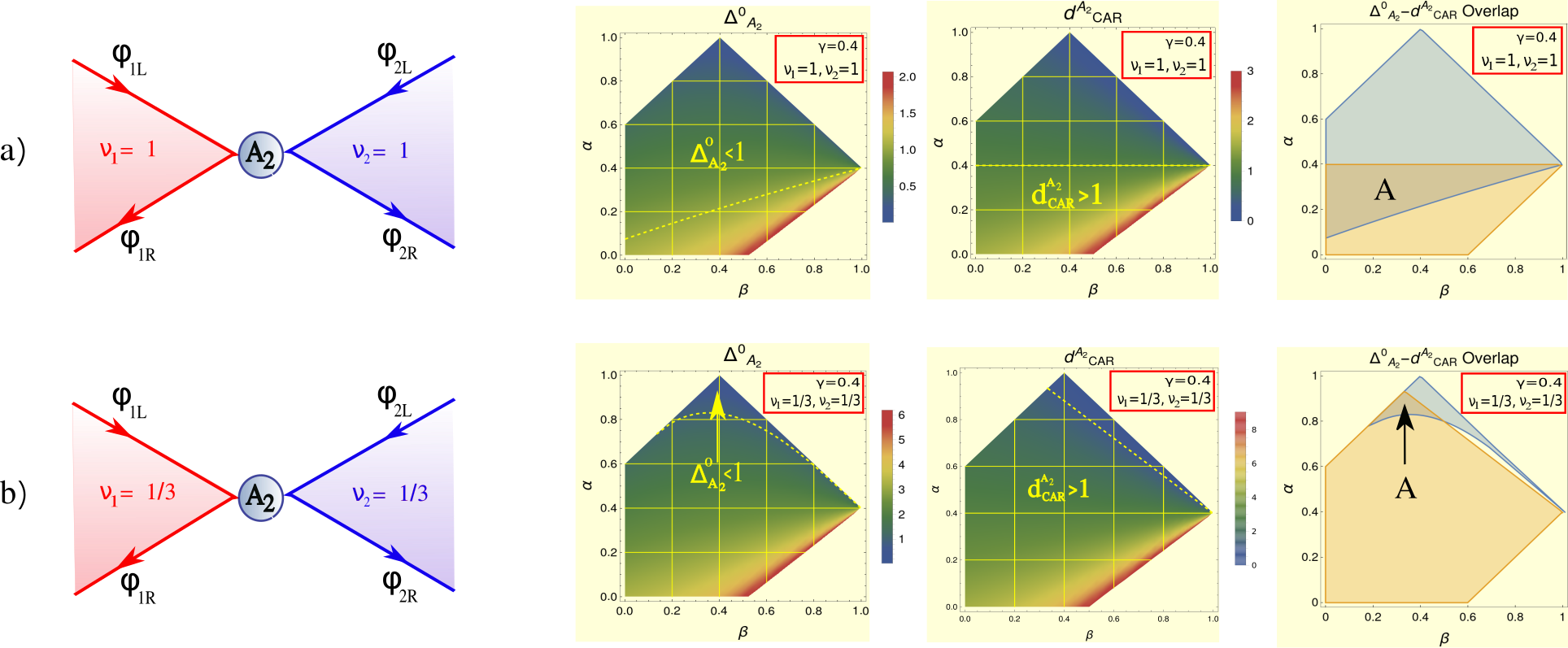}
    \caption{The schematic figure on the left shows the unfolded version of the junction of bilayer QH edge states with equal filling fraction, $ \nu_{1}=\nu_{2} \in \lbrace 1,1/3\rbrace$,  tuned to $\mathrm{A_{2}}$ fixed point. Fig. a) shows the junction of edge states with $\nu_{1}=\nu_{2}=1$. The three plots correspond to $\Delta^{0}_{A_{2}}$, $d^{A_{2}}_{CAR}$ and '$\Delta^{0}_{A_{2}}-d^{A_{2}}_{CAR}$ overlap' plot showing the interaction parameter region for which TDOS is enhanced and junction is stable against CAR operator for $\gamma = 0.4$. Fig. b) shows the junction of edge states with $\nu_{1}=\nu_{2}=1/3$. The three plots correspond to $\Delta^{0}_{A_{2}}$, $d^{A_{2}}_{CAR}$ and '$\Delta^{0}_{A_{2}}-d^{A_{2}}_{CAR}$ overlap' plot showing the interaction parameter region for which TDOS is enhanced and junction is stable against AR operator for $\gamma = 0.4$. Region 'A' in $\Delta^{0}_{A_{2}}-d^{A_{2}}_{CAR}$ overlap plot, in both fig. a) and b), denotes the interaction parameters for which $\Delta^{0}_{A_{2}} < 1$ and $d^{A_{2}}_{CAR} > 1$ simultaneously. When the junction is tuned to $\mathrm{CA_{2}}$ fixed point, we can speculate the TDOS enhancement and stability (against AR operator) scenarios through the symmetry relations between the $\mathrm{A_{2}}$ and $\mathrm{CA_{2}}$ fixed point in the $\alpha\leftrightarrow \gamma$ exchange.}
    \label{fig: nu1_eq_nu2_A2_andreev_stability}
\end{figure*}

\end{document}